\documentclass[nofootinbib,twocolumn,longbibliography]{revtex4-1}
\usepackage{graphicx}

\usepackage{amsmath}
\usepackage{amstext}
\usepackage{amssymb}

\usepackage{xfrac}
\usepackage[colorlinks,citecolor=blue]{hyperref}
\usepackage{graphicx}
\usepackage{amsmath}
\usepackage{amstext}
\usepackage{amssymb}
\usepackage{amsfonts}
\usepackage{longtable,booktabs}
\usepackage{hyperref}\usepackage{url}
\usepackage{subfigure}%
\usepackage{dsfont}
\usepackage{soul} 
\usepackage{amsbsy}
\usepackage{dcolumn}
\usepackage{amsthm}
\usepackage{bm}
\usepackage{esint}
\usepackage{multirow}
\usepackage{hyperref}
\hypersetup{
    colorlinks=true,
    linkcolor=blue,
    filecolor=magenta,
    urlcolor=blue,
}
\usepackage{cleveref}

\usepackage{mathrsfs}
\usepackage{amsfonts}
\usepackage{amsbsy}
\usepackage{dcolumn}
\usepackage{bm}
\usepackage{multirow}
\usepackage{color}
\usepackage{pifont}

\newcommand{\ket}[1]{|#1\rangle}

\newcommand{\comments}[1]{}

\def\Z{\mathbb{Z}}

  \usepackage{extarrows}
 
\usepackage{datetime}

\usepackage{comment}

\usepackage{lineno}

\begin{document}


 \title{Exactly Solvable Fracton Models for  Spatially Extended Excitations}
 \title{Fracton physics of spatially extended   excitations}


\author{Meng-Yuan Li}
\affiliation{School of Physics, Sun Yat-sen University, Guangzhou, 510275, China}

\author{Peng Ye}\email{yepeng5@mail.sysu.edu.cn}
\affiliation{School of Physics, Sun Yat-sen University, Guangzhou, 510275, China} 


\date{\today}

\begin{abstract}
 		
Fracton topological order hosts fractionalized point-like excitations (e.g., fractons) that have restricted mobility. In this article, we explore even more bizarre realization of fracton phases that admit  spatially extended excitations with restriction on both mobility and deformability.    First, we   present  exactly solvable lattice quantum frustrated spin models and study their ground states and excited states analytically.  We construct a family tree in which parent models and descendent models  share   excitation DNA.  Second, with the help of solvability and novel excitation spectrum of these models, we  initiate the first-step of general discussions on quantitative and qualitative properties of spatially extended excitations  whose   mobility and deformability are restricted to some extent.   Especially, as a useful viewpoint for understanding such fracton-physics, all excitations are divided into four mutually distinct sectors, namely, simple excitations, complex excitations, intrinsically disconnected excitations, and trivial excitations. Several implications in, e.g., condensed matter physics and  gravity are briefly discussed.   

%
%
%
	 
\end{abstract}



\maketitle
\tableofcontents


\section{Introduction}
\label{sec:intro}
 
%
\subsection{Mobility and deformability}

  \begin{figure*}[t]
	\centering  
	\includegraphics[width=0.98\textwidth]{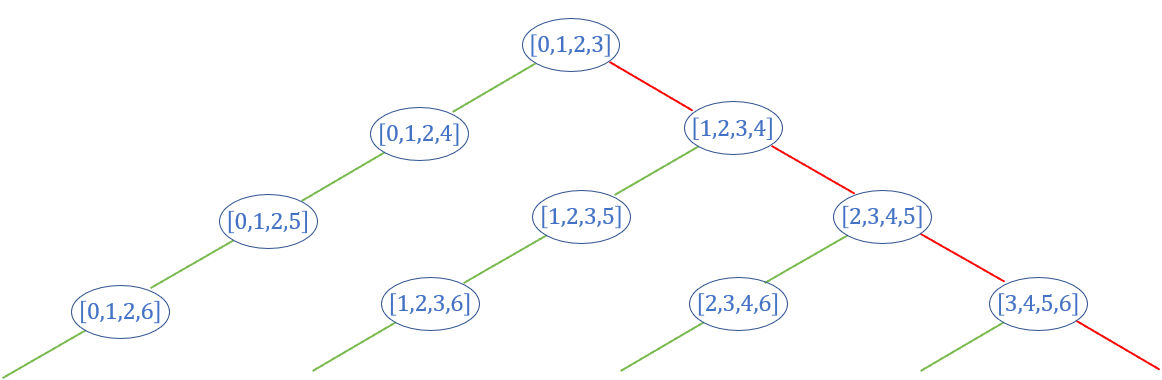}
	\caption{   The family tree of exactly solvable fracton models in all dimensions. The models are labeled by $4$ dimension indexes, which will be introduced in Sec.~\ref{subsec:cons_family}. Considering that a model may share a lot of similarities with a lower dimensional model, a part of exactly solvable models can be organized in a tree diagram. Here every straight line (in either green or red) links a low dimensional parent model and a high dimensional descendent model. In general, a descendant model may share some ``excitation DNA'' with its parent model. For example, a red line means the $(i,i+1)$-type excitations in the parent model are promoted to $(i+1,i+2)$-type excitations in the descendent model (as we shall exemplify in Table.~\ref{table:excitations}). Even though, due to the existence of $\mathsf{E}^c$ excitations, the spectrum of a model can still be quite unpredictable, as we shall demonstrate in the following sections.}
	\label{fig_tree}
\end{figure*} 

Hunting for unconventional orders---beyond symmetry-breaking orders---is one of missions of  modern condensed matter physicists. One of popular examples of unconventional orders is topological order that  supports nontrivial topological excitations, e.g., anyons in the fractional quantum Hall effect  \cite{wen_stacking,wen1990topological}.  The creation and annihilation quantum  operator of a single topological excitation in the bulk, e.g., $e$ particle in toric code model, must be nonlocal.   This nonlocality also leads to   robustness of topological order against local perturbations, partially forming the argument on robustness of topological quantum computation\cite{sarma_08_TQC}.   Meanwhile, proper local operators can be constructed to  spatially move a topological excitation. Even in a tight-binding model,   a ``topologically trivial'' electron can locally hop   from site-$j$ to site-$i$ under the local operator $\sim c^\dagger_i c_j$. This property is ``free mobility''.    
Recently, the invalidity of this seemingly obvious property  has been found   in a class of   many-body systems that support topological excitations called ``fractons''  \cite{Chamon05,Vijay2015}.        If one tries to move a fracton,  extra fractons will be created nearby, causing  unfavorable energy penalty.  Quantum mechanically, the mobility restriction  is deeply rooted in  the lack of local operators that   act on a one-fracton excited state by \emph{merely} changing the fracton's location. There are two issues about lack-of-mobility to be clarified. First, in some sense, this lack of mobility in absence of external disorders and impurities is more or less similar to   the concept of  self-localization phenomenon although the latter arises in some other strongly-correlated systems, e.g., in Ref.  \cite{ye_wang},  via   very different microscopic origins. Second, lack of mobility leads to flat energy dispersion relation, but flat energy dispersion relation is insufficient for defining a \emph{fracton phase}. All excitations in the $\Z_2$ topological order fixed-point ``toric code model''   have flat dispersion but all excitations can be locally moved.

Surprisingly, such   stringent restriction on mobility  did not trivialize underlying physics at all. On the contrary, it has been discovered that   mobility restriction leads to unexpectedly rich quantum phenomena of many-body physics, dubbed ``\emph{fracton-physics}''. For example,  in some exactly solvable models,  ground state degeneracy (GSD) is not only topological but also dependent on the system size! More specifically, GSD of some models  \cite{Shirley2018}  may grow exponentially    with respect to the length/width/height  of 3D systems, while,  mutually orthogonal degenerate ground states  are   strictly indistinguishable under  any local measurements.  Generally speaking,  such many-body systems possess a part of intrinsic defining properties of \emph{pure topological order}\footnote{Hereafter, for avoiding confusion of terminologies, we use   ``pure topological order''  to denote the  well-known concept ``topological order'' \cite{wen1990topological,wen_stacking}. \label{footnote_pure_topo}}  but the    thermodynamical limit of these systems turns out to be  quite subtle and unusual.     Such  an ``unconventional'' type of topological order, dubbed  ``\emph{fracton topological order}'' represents a brand-new line of thinking about strongly-correlated topological phases of matter, and has been gaining  much attention recently. Researchers have successfully made connection betwen fracton-physics and vast subfields of   theoretical physics, including glassy dynamics, foliation theory, elasticity, dipole algebra, higher-rank global symmetry, many-body localization, stabilizer codes, duality, gravity, quantum spin liquid, and higher-rank gauge  theory \cite{Vijay2015,Vijay2016,Prem2017,Chamon05,Vijay2015,Shirley2019,Ma2017,Haah2011,Bulmash2019,Prem2019,Bulmash2018,Tian2018,You2018,Ma2018,Slagle2017,Halasz2017,Tian2019,Shirley2019b,Shirley2018a,Slagle2019a,Shirley2018,Prem2017,Prem2018,Pai2019,Pai2019a,Sala2019,Kumar2018,Pretko2018,Pretko2017,Ma2018,Pretko2017a,Radzihovsky2019,Dua2019,PhysRevLett.122.076403,haahthesis,PhysRevX.9.031035,2019arXiv190411530Y,2019arXiv190913879W,pretko18string,pretko18localization,PhysRevB.100.125150,PhysRevB.99.245135,PhysRevB.97.144106,PhysRevB.99.155118}.  
Some of these subfields,  from previous points of view, seem no doubt ``orthogonal'' to each other!  Being topically  related to the present article, exactly solvable lattice models  in the literature  (e.g.,  \cite{Vijay2015,Shirley2019,Vijay2016a,Ma2017,Prem2018,Haah2011,Bulmash2019,Prem2019,Bulmash2018,Tian2018,You2018,Ma2018,Slagle2017,Halasz2017,Tian2019})   have been reported  in 3D lattice quantum frustrated spin models of  type-I and type-II. In type-I series, e.g, the X-cube model  \cite{Vijay2015}, the low-lying excitation spectrum supports both  fractons   and  ``subdimensional particles'' whose mobility is free only inside a subspace (e.g., straight lines formed by links and flat plane formed by faces of dual lattice) of 3D cubic lattice.  On the contrary, in type-II series, e.g., the Haah's code  \cite{Haah2011}, all topological excitations are fractons.  For readers who are interested but unfamiliar with the rapid progress in the field of fracton-physics, an up-to-date   review   in  Ref.  \cite{2020arXiv200101722P}  is recommended.

Currently,  the main stream on the topic of fracton topological order focuses  on particle excitations which are point-like\footnote{Just like pure topological order, the geometric shapes of excitations   can be properly defined by using continuous spacetime only after smoothing lattice. For example, $m$ particle  in the toric code model on a square lattice is actually labeled by a plaquette operator whose eigenvalue is flipped, but can be regarded as a point-like object once the background lattice is smoothen. It also means that all geometric structures below the lattice spacing are invisible.\label{ft_smooth}}.  Nevertheless,  in addition to particles, being a striking theoretical progress  in condensed matter physics, \emph{spatially extended excitations}, e.g., string and  membrane excitations, have been  systematically constructed in 3D and higher dimensional pure topological order  \cite{string2}.  It should be kept in mind that, higher-dimensional pure topological order    has been analyzed towards a unified mathematical framework     \cite{lantian3dto1,lantian3dto2,string8}. In the presence of spatially extended excitations,  plentiful quantum phenomena and microscopic justification have been reported analytically, such as   exotic entanglement, symmetry enrichment, adiabatic braiding statistics and topological quantum field theory, and higher-category  \cite{lantian3dto1,lantian3dto2,yp18prl,ypdw,wang_levin1,jian_qi_14,string5,PhysRevX.6.021015,string6,ye16a,YeGu2015,corbodism3,YW13a,ye16_set,2018arXiv180101638N,2016arXiv161008645Y,string4,PhysRevLett.114.031601,3loop_ryu,string10,2016arXiv161209298P,Ye:2017aa,Tiwari:2016aa,2012FrPhy...7..150W}.  \textit{Therefore,  it is  quite  valuable to move forward to explore underlying physics of  spatially extended excitations that  cannot freely   {move} and  {deform}}.  With the preparation and   interests from both pure topological order side and fracton-physics side,  it is time to make efforts to  study the highly unexplored  marriage of spatially extended excitations and mobility/deformability restriction. Despite less progress compared to particle excitations,  to the best of our knowledge, there has been one intriguing  field-theoretical analysis  on ``\emph{fractonic lines}'' in Ref.~\cite{pretko18string}, i.e., completely immobile strings.  It was claimed that, the presence of such exotic excitations is tightly related to sophisticated higher-rank gauge theory and  potentially beneficial to  quantum error-correction and quantum storage.

The main results of   this article can be summarized in two aspects. \emph{First, we   present  exactly solvable lattice quantum frustrated spin models in three and higher dimensions $D\geq 3$.}  All models   reduce  to the aforementioned  X-cube model once dimensions are lowered to 3D. But for the definite space dimensions higher than 3D, there are more than one models. All models  form  a hierarchical structure of model Hamiltonians. Some representative series of models are   illustrated in Fig.~\ref{fig_tree} since \textit{a part of} excitations in these models obey simple dimension reduction rules.  In these   models, topologically excited states contain not only   fractonic strings \cite{pretko18string}, but also more complex variaties as to be discussed in the main text.  \emph{Second, motivated by  these  models, we  initiate the first-step of general discussions on spatially extended excitations  whose   mobility and deformability are restricted to some extent. Both qualitative and quantitative properties in such exotic fracton physics will be discussed.}  Along this line of thinking,  one challenging problem   is to characterize and  classify topological excitations based on    mobility and deformability against local operators. Pictorially, a spatially extended object in classical mechanics may leave  its original position   via either \emph{rigid translation} or \emph{elastic deformation}. To measure the ability of realizing these two processes, we introduce respectively ``mobility'' and ``deformability'' as mentioned above.  Such a classical scenario looks more complicated than point-like excitations, which motivates us to carefully examine how excitations are deformed and moved under local quantum operators.

    {Before moving on, we  provide some justification for models  in dimensions higher than the physically relevant dimensions of   three.} Traditionally, it is meaningful to study condensed matter systems only in dimensions of one, two, and three. It seems unreasonable to go beyond.   Nevertheless,  research in all dimensions has been very common, especially in the field of topological phases of matter.  Organizational principles or mathematical structures of topological phases of matter are often unveiled during the systematic treatment  via varying  dimensions.  For example, the periodic table of free-fermion gapped states with symmetry shows interesting periodic behaviors when increasing space dimensions topological insulators, by the procedure of dimensional reduction \cite{kitaev_period,PhysRevB.78.195125,Qi2008,Ryu_2010}.   In the group-cohomology construction of bosonic symmetry-protected topological phases (SPT), SPTs in higher dimensions can be constructed by SPTs in lower dimensions via    K\"unneth formula of cohomology theory\cite{Chenlong,Chen:2014aa}, which is physically discussed via the ``decoration scenario''.  Also in SPTs, there exist exotic dimension-dependent patterns for general response theories of   bosonic integer quantum Hall states in all even (spatial) dimensions and bosonic topological insulators in all odd (spatial) dimensions \cite{lapa17}. There are also interesting discussions on anomalous topological phases of matter in 3D \cite{PhysRevD.92.085024,2016arXiv161008645Y,ye16_set,ye18a}. Quantum anomaly of these states is expected to be   canceled by 4D bulk states. \cite{JianXu2018} studies how to realize interacting topological insulators with synthetic dimensions and \cite{2019arXiv191205565F} proposes a 4D exactly solvable  SPT model  beyond cohomology. The last example is   the general theory of topological order in all dimensions \cite{string8} which has also been mentioned above.

        \subsection{Sectorization of Hilbert space}\label{sec_intro_sector}
 
\begin{figure}[t]
\centering
\includegraphics[width=6.5cm]{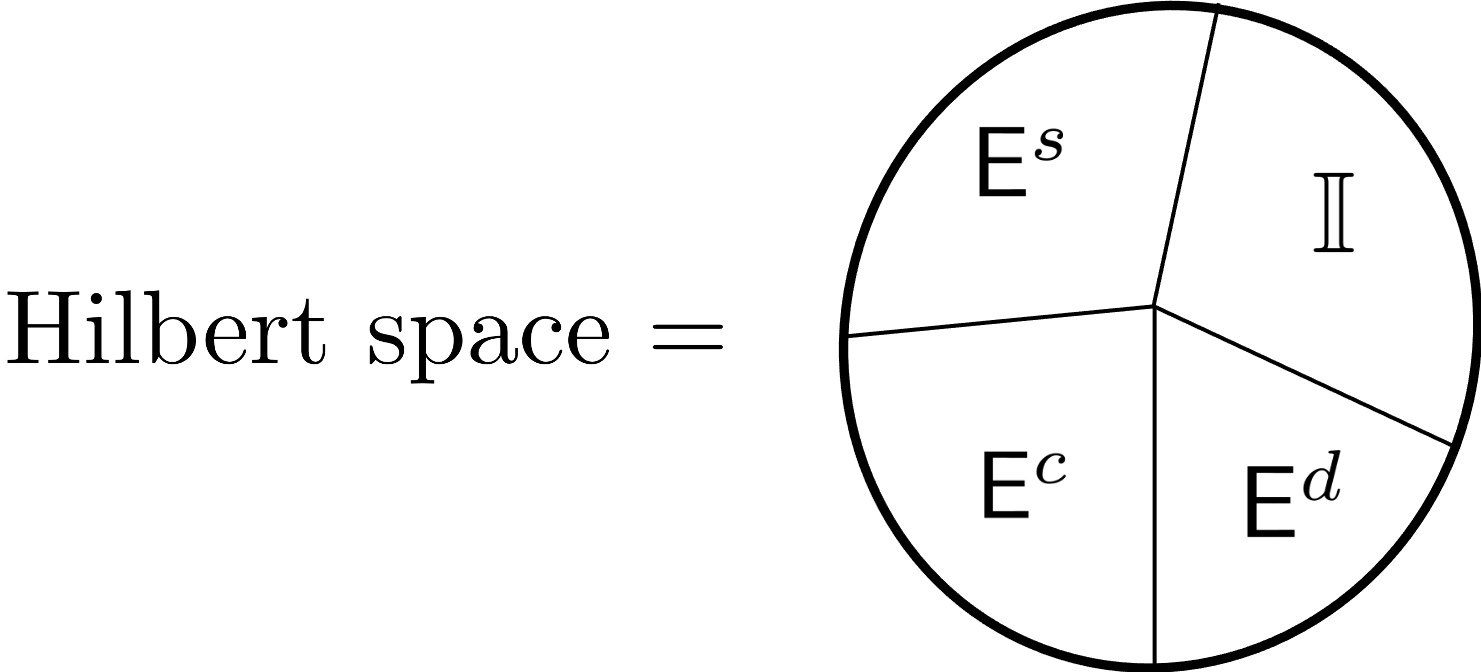}
\caption{Sectorization of Hilbert space.  It is guaranteed that   two excitations belonging to two different sectors definitely cannot be changed to each other by any local operators.}\label{figure_hilbertspace}
\end{figure}

Among many properties of spatially extended excitations, in this article, we focus on  mobility and deformability under local operators. In the long-wavelength limit, we demand that the size of spatially extended excitations is sufficiently large compared to correlation length. All local operators are supported in the space whose size is much smaller than the excitation size such that topology of configuration space in the presence of  excitations (e.g., defects) keeps unaltered under local operators.  We find that the Hilbert space of models that support fracton topological order can be divided into four sectors, as shown in Fig.~\ref{figure_hilbertspace}. $\mathbb{I}$ incooporates all trivial excited states e.g., local spin flipping, including ground states themselves as well. Trivial excitations can be created by local operators above the ground states.  The remaining three sectors are three mutually  distinct classes of topological excitations:  \textit{simple excitations} $\mathsf{E}^s$, \textit{complex excitations} $\mathsf{E}^c$ and \textit{(intrinsically) disconnected excitations} $\mathsf{E}^d$.   Below we shall define  $\mathsf{E}^s,\mathsf{E}^c,\mathsf{E}^d$.  

Let us first assume geometric shape of excitations is connected. At infrared scales where lattice has been smoothen (see footnote \ref{ft_smooth}), simple excitations denoted by $\mathsf{E}^s$ have manifold-like shape, e.g., point-like, string-like, membrane-like.  All   these excitations, once withdrawing the restriction on mobility and deformability, can appear in   pure topological order  \cite{lantian3dto1,lantian3dto2}.  Mathematically, all these geometric structures can be locally regarded  as a $n$-dimensional Euclidean space   (see Page 219 of Ref.~ \cite{EGUCHI1980213}), where $n=0$ for points (i.e., particles), $n=1$ for strings, $n=2$ for membranes, $\cdots$.   In order to   characterize  simple excitations in a unified framework,  in the present article,   we introduce a pair of  integers $(n,m)$.  Here, $m$ denotes the dimension of the subspace where excitations can freely move and deform\footnote{Obviously, $m$ is insufficient to  uniquely label a general subspace. For example,   it is reasonable to consider a model where a string excitation  is  movable and deformable inside a certain solid torus. Nevertheless, we only focus on the simplest situation: for dimension-$m$, the subspace is a stacking of infinite parallel \emph{straight} lines ($m=1$), \emph{flat} planes ($m=2$), $\cdots$.}. Therefore,  fractons  are simply  labeled by $(0,0)$.   Likewise, a string excitation whose  mobility and deformability   are allowed within a plane is a $(1,2)$-type excitation.  Obviously, if $m=D$, then such excitations are actually mobile and deformable  in the whole $D$-dimensional space, which   exactly covers all excitations  in pure topological order.

  {On the other hand}, for complex excitations denoted by $\mathsf{E}^c$, physical properties of both geometric shapes and mobility  are entirely different from the above description of simple excitations.  The physical characterization (e.g.,  creation operators, excitation energy, fusion rules,  mobility) is far more intricate than that of simple excitations. When we consider the connected configurations of excitations, the shapes of complex excitations can only be non-manifold-like  \cite{EGUCHI1980213},  so a pair of number $(n,m)$ is no longer a good label. And consequently, the description of mobility and deformability becomes more complicated.  But it is definite that any point-like excitations cannot belong to $\mathsf{E}^c$ since a point is always  a manifold. The first example of  $\mathsf{E}^c$ that we will introduce in the main text is dubbed ``chairon'' due to its ``legless chair-like'' shape  as shown in Fig.~\ref{fig_3b}. More complicated examples such as ``yuons'', ``xuons'' and ``cloverions'' will be discussed in the main text associated with   Fig.~\ref{fig_com_yuon} and Fig.~\ref{fig_com_xuon}.

Simple excitations and complex excitations defined above are restricted to geometrically connected shapes. Nevertheless, we should naturally generalize these definitions in order to incorporate some excited states with disconnected pieces.   For disconnected shapes, excitations (more precisely ``excited states'') can be divided into two subclasses: $\mathsf{E}^d$ and $\widetilde{\mathsf{E}}^{d}$. The shapes  of all excitations in $\mathsf{E}^d$ are said to be ``intrinsically disconnected'':   there is no way to fuse disconnected pieces to connected shapes due to restrictions of mobility and deformability. To some extent, the existence of $\mathsf{E}^d$ is a hallmark of fracton topological order.  On the contrary, all excitations in $\widetilde{\mathsf{E}}^{d}$ can always be fused into excitations with connected shapes of either manifold-like or non-manifold-like\footnote{In this article, we only consider the simplest fusion process: output channel is unique.}. In this sense, all excitations in $\widetilde{\mathsf{E}}^d$ can be fully covered by either $\mathbb{I}$, $\mathsf{E}^s$ or $\mathsf{E}^c$.  If non-manifold-like shape is the \emph{only} option of fusions, the excitation with disconnected shape is said to be in $\mathsf{E}^c$ sector.    {In short, we do not separately consider $\widetilde{\mathsf{E}}^d$ in Fig.~\ref{figure_hilbertspace}.}  Applying this sectorization of Hilbert space to the three-dimensional X-cube model can be found in Sec.~\ref{subsec:rev_of_xcube}.

%

In summary, the Hilbert space (eigenstate spectrum) can be divided into four sectors. In a certain sector, all excitations,  after  being moved and deformed by arbitrary local operators,  always stay inside the sector.    In other words, it is impossible to change one excitation in a given sector, via local operators, to another excitation in another sector.    We will see in this article, this sectorization scheme is very useful in fracton-physics of spatially extended excitations.

\subsection{Exactly solvable   models}

In the main text of this article,   we discuss the fracton physics of spatially extended excitations through exactly solvable models.    
Since there are some additional tunable degrees of freedom  within the same spatial dimension $D$, we find that at least four dimension indices (note: $D$ is included) are necessary to label  a model. In the following part of this article, we will use a tuple $[d_n,d_s,d_l, D]$ with a series of constraints required by exact solvability conditions. Technical details of each integer will be given  in  the main text.    For $D=3$,  $[0,1,2,3]$ is the only model that is exact solvable. In fact, this 3D model is nothing but the  standard X-cube model  \cite{Vijay2016}. 

  While there are usually more than one models for a more general  $D$, we first pick a typical model series---$[D-3,D-2,D-1,D]$---to systematically unveil exotic properties of spatially extended excitations with restricted mobility and deformability.  The model-$[0,1,2,4]$  has similar simple excitation contents as 3D X-cube model. But the model-$[1,2,3,4]$ supports a very fruitful  topological excitation spectrum with all three non-trivial sectors, which will be studied in details in the main text.  Some other models, such as $[1,2,3,5]$ will also be studied in which chairons, xuons and cloverions are found. We finally provide a family tree in Fig.~\ref{fig_tree} to summarize some models that share similar properties of excitations. Several interesting rules are found and summarized as a family tree according to the relation of excitation spectrum between parent models and descendent models (see the caption for details).

  Some examples   are summarized in Table~\ref{table_all_excitations}.  It is obvious from the table that pure topological order only supports topological excitations in $\mathsf{E}^s$ with the label $(n,m)=(n,D)$. For example, in pure topological order represented by 3D toric code model, point-like and string-like excitations are labeled by $(0,3)$ and $(1,3)$ respectively, both of which belong to $\mathsf{E}^s$ sector.   
  On the other hand, fracton topological order support more than that. $X$-cube model supports  topological excitations of $\mathsf{E}^s$ and $\mathsf{E}^d$ sectors   while some models constructed in this article support all possible sectors.

\begin{table*}[htp] 	
	\caption{Excitations sectors of different models.   More details about the excitations in the listed models are respectively given in Table.~\ref{table:1234_excitations}, Table.~\ref{table:0124_excitations}, Table.~\ref{table:1235_excitations} and Table.~\ref{table:5d_excitations}. $\checkmark$ and $\times$ denote existence and nonexistence respectively.}
	
\begin{tabular}{cc cc c c}
		\hline
		
		\hline
		Exactly solvable    models    & \begin{minipage}{0.3in}$\mathbb{I}$ \end{minipage}& \begin{minipage}{0.3in}$\mathsf{E}^s$ \end{minipage}& \begin{minipage}{0.3in} $\mathsf{E}^d$ \end{minipage}& \begin{minipage}{0.3in} $\mathsf{E}^c$\end{minipage} \\
		\hline
		\begin{minipage}{2in} SPT (e.g., cohomological models)   \end{minipage}& $\checkmark$ & $\times$ & $\times$& $\times$
		\\
		\begin{minipage}{3in}	Pure topological orders (e.g., 3D toric code models)\end{minipage}& $\checkmark$ & $\checkmark$ & $\times$& $\times$
		\\

		X-cube model ($[0,1,2,3]$) & $\checkmark$& $\checkmark$ & $\checkmark$ & $\times$
		\\
		$[1,2,3,4]$& $\checkmark$ & $\checkmark$ & $\checkmark$ & $\checkmark$
		\\
		$[0,1,2,4]$& $\checkmark$ & $\checkmark$ & $\checkmark$ &  $\times$
	    \\
	    $[1,2,3,5]$& $\checkmark$ & $\checkmark$ & $\checkmark$ &  $\checkmark$
	    \\
	    $[0,1,4,5]$& $\checkmark$ & $\checkmark$ & $\checkmark$ & $\times$
	    \\
	    $[0,2,4,5]$& $\checkmark$ & $\checkmark$ & $\checkmark$ & $\times$
	    \\
	    $[0,3,4,5]$& $\checkmark$ & $\checkmark$ & $\checkmark$ & $\times$
	    \\
	    $[2,3,4,5]$& $\checkmark$ & $\checkmark$ & $\checkmark$ & $\checkmark$
	    \\
		\hline
		
		\hline	
	\end{tabular}
	\label{table_all_excitations}
\end{table*}

\subsection{Outline}
 
The remainder of this article is organized as follows.   In Sec.~\ref{sec:rev_not}, we introduce geometric notations that are necessary to symbolize derivations in hyper-cubic lattices. With the help of these notations, all derivations are transformed into a computable algebraic way.  

 In Sec.~\ref{sec:1234series}, we  discuss a series of models called ``$[D-3,D-2,D-1,D]$ models''. A general introduction of the series is given in Sec.~\ref{subsec:cons_1234}, while the X-cube model introduced in Sec.~\ref{subsec:rev_of_xcube} has been naturally incorporated in this series and labeled by $[0,1,2,3]$.   For   beginners of fracton-physics, it is highly recommended to go through the X-cube model where some notations and physics are useful for   later discussions. In Sec.~\ref{subsec:ex_1234} and Sec.~\ref{subsec:com_ex_1234}, we work out the model-$[1,2,3,4]$ that exemplifies the construction of spatially extended excitations of both ``simple'' and ``complex'' categories. In this model, simple excitations are composed by fractons labeled by $(0,0)$, volumeons labeled by $(0,3)$, and strings labeled by $(1,2)$ with 6 flavors. Complex excitations of this model are chairons and yuons. Considering that in pure topological orders excited states with separated loops are rarely discussed, Sec.~\ref{sec_multiple_loop} is devoted to a detailed discussion of such states, as now they are of great importance to understand the bizarre behavior of $[D-3,D-2,D-1,D]$ models.  The construction of general $(i,i+1)$-type excitations and its possible relationship with gravity is also presented in Sec.~\ref{subsec:ex_1234series}.

  In Sec.~\ref{sec:family}, we present a general procedure to produce a whole class of exactly solvable lattice models for fracton topological order in all dimensions $D\geq3$. Each model is labeled by four integers $[d_n,d_s,d_l, D]$, which means that the above model series $[D-3,D-2,D-1,D]$ is just a tip of iceberg of model family. The construction and general discussion of the whole model family is presented in Sec.~\ref{subsec:cons_family}, and a family tree based on similarity of excitation spectrum is drawn in Fig.~\ref{fig_tree} in Sec.~\ref{subsec:tree}. In Sec.~\ref{subsec:fractonic} and Sec.~\ref{subsec:com_1235} we concretely discuss the model-$[1,2,3,5]$, while the model-$[0,1,2,4]$ is also discussed briefly in Sec.~\ref{subsec:fractonic}. Many examples of complex excitations, like $\beta$-chairons, cloverions and xuons are analyzed in details. 
  
 Sec.~\ref{sec:conclusion} is devoted to concluding remarks. Several related problems are presented for the future investigation.

\section{Preliminaries of geometric notations}
\label{sec:rev_not}

 \subsection{Coordinate system and definition of $n$-cube}
\begin{figure}[t]
\centering
\includegraphics[scale=0.33]{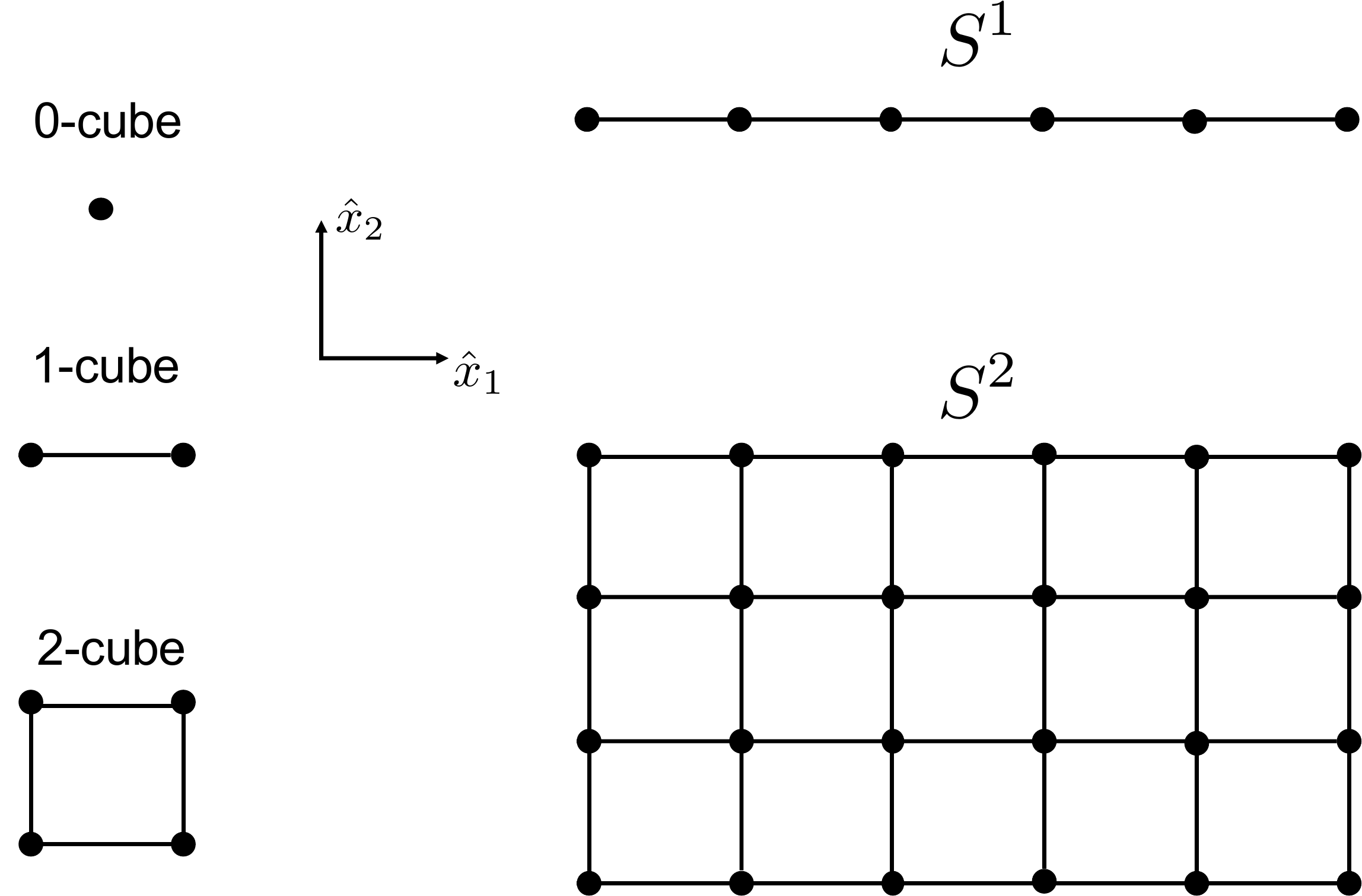}
\caption{Some examples of   geometric objects $n$-cube and $S^j$ embedded in $D$-dimensional hypercubic lattice. }\label{figure_cube}
\end{figure}

In this article, we're mainly interested in the high dimensional models, so it's highly desirable   to define and unify a group of notations for describing high dimensional objects. First, we introduce ``$n$-cube''. It is $n$-dimensional analog of a common ``cube'', and we use the symbol $\gamma_n$ to denote an $n$-cube. Some simple examples are shown in Fig.~\ref{figure_cube}. In other words, $0$-cube, $1$-cube and $2$-cube are respectively a lattice site, a link and a plaquette. Without loss of generality, we   set the hypercubic lattice with periodic boundary condition to be $D$-dimensional with lattice constant $a=1$. Therefore, we can refer to every $n$-cube in the lattice by a unique Cartisian coordinate, \textit{which is the coordinate of the geometric center of  $\gamma_n$}. Obviously, the coordinate representation of $\gamma_n$ is composed by $n$ half-integers and $(D-n)$ integers. For example, a usual vertex is a $0$-cube. In the remainder of this article, we may simply use the coordinate of an $n$-cube to refer to the $n$-cube itself, since the coordinate can uniquely label an $n$-cube.   

In $D$-dimensional lattice, there are $D$ orthogonal directions: $\hat{x}_1,\hat{x}_2,\cdots,\hat{x}_D$. For a specific  $\gamma_{d_n}=(x_1,x_2,\dots,x_D)$,  the set $\mathcal{C}^i_{\gamma_{d_n}}$ is a collection of $(D-d_n)$ orthogonal directions along which the coordinates of $\gamma_{d_n}$ are integer-valued.  Likewise,  the set $\mathcal{C}^h_{\gamma_{d_n}}$ is composed by   $d_n$ directions along which   $\gamma_{d_n}$ has half-integer coordinates. For example, in 3D cubic lattice, for plaquette (i.e. $2$-cube) $\gamma_2=(\frac{1}{2},\frac{1}{2},0)$, we have $\mathcal{C}^i_{\gamma_2}=\{\hat{x}_3\}$ and $\mathcal{C}^h_{\gamma_2}=\{\hat{x}_1,\hat{x}_2\}$.

\subsection{Leaf spaces associated with a given  cube}\label{sec_pre_leaf}

In addition to the notion of $n$-cube, we     also   introduce a useful subspace, namely, $d_l$-dimensional leaf space associated with a given $\gamma_{d_n}$.  By ``associated'', we mean that  the $\gamma_{d_n}$ must be  fully embedded in  the leaf space $l$, and $d_l>d_n$ is assumed.   Symbolically, we use  $l=\langle\hat{x}_{i_1} ,\hat{x}_{i_2} ,\cdots,\hat{x}_{i_{d_l}}\rangle$ to \textit{uniquely} denote such a subspace. Among these $d_l$  orthogonal directions,    $d_n$ ones come from the set $\mathcal{C}^h_{\gamma_{d_n}}$.   As a result, the remaining $(d_l-d_n)$ ones are arbitrarily picked  from the set $\mathcal{C}^i_{\gamma_{d_n}}$. Therefore, there are combinatorially $\binom{D-d_n}{d_l-d_n}\equiv\frac{(D-d_n)!}{(d_l-d_n)!(D-d_l)!}$ different choices of leaf space associated with the given $\gamma_{d_n}$.  

It must be noted that, any lattice site inside the leaf space $l$   has a coordinate with $D$ components since the lattice site is in fact a point in $D$-dimensional lattice.  Among $D$ components, $d_l$ components are free variables with orthogonal directions $\hat{x}_{i_1} ,\hat{x}_{i_2} ,\cdots,\hat{x}_{i_{d_l}}$, which spans a $d_l$-dimensional subspace. The remaining $(D-d_l)$ coordinate components are fixed and simply   equivalent to corresponding coordinate components of $\gamma_{d_n}$. Therefore, a leaf associated with a given $\gamma_n$ can be uniquely labeled by $l$  as long as $\gamma_{d_n}$ is specified.  For such a leaf $l$, we can define a set of orthogonal directions $\mathcal{L}=\{\hat{x}_{i_1} ,\hat{x}_{i_2} ,\cdots,\hat{x}_{i_{d_l}}\}$, which will be used later.


Let us apply the above notation to the X-cube model (to be introduced in Sec.~\ref{subsec:rev_of_xcube}).  The X-cube model has foliation structure  \cite{Shirley2018a,Shirley2018,Shirley2019b,Slagle2019a,Prem2018}, where the leaf space dimension $d_l=2$, and the model dimension $D=3$. The direction index $i$ in Eq.~(\ref{eq:-3-2-10Hamiltonian}) can be also seen as an index for a leaf space $l$.  For example,  when $i=x$ and the vertex is $(0,0,0)$ (i.e., a $0$-cube), $B^x_{(0,0,0)}$ corresponds to  the nearest four $\sigma^z$'s inside the $\langle\hat{y},\hat{z}\rangle$ leaf (i.e., $\hat{y}-\hat{z}$ plane with $x=0$). 
In this manner, for the Hamiltonian in the form of 
\begin{align}
H_{X-cube} = -J\sum_{\{\gamma_3\}} A_{\gamma_3} - K \sum_{\{\gamma_0\}} \sum_{l} B^l_{\gamma_0}\,  
\label{equivalent_form_xcube}
\end{align} 
which is in fact the standard X-cube model that will be given in Eq.~(\ref{eq:-3-2-10Hamiltonian}),  
there are in total $\binom{3-0}{2-0}=\binom{3}{2}=3$ different leaf spaces: $\langle\hat{y},\hat{z}\rangle\,,\langle\hat{x},\hat{z}\rangle\,,\langle\hat{x},\hat{y}\rangle$ planes associated with the vertex $(0,0,0)$. All of these planes pass through the vertex $(0,0,0)$.  

Besides, as higher dimensional leaf spaces shall be used in the following sections, here it's also beneficial to give some examples of leaves in high dimensional models:

\textit{Example 1.---} In the model-$[1,2,3,4]$ (to be studied in Sec.~\ref{subsec:ex_1234}),  the total space dimension $D=4$ and the leaf space dimension $d_l=3$. For a $1$-cube $\gamma_1=(0,0,0,\frac{1}{2})$, there are $\binom{4-1}{3-1}=3$ leaves associated with it, which are respectively $\langle\hat{x}_1,\hat{x}_2,\hat{x}_4\rangle$, $\langle\hat{x}_1,\hat{x}_3,\hat{x}_4\rangle$ and $\langle\hat{x}_2,\hat{x}_3,\hat{x}_4\rangle$. The coordinate component $x_3$ of each lattice site inside $\langle\hat{x}_1,\hat{x}_2,\hat{x}_4\rangle$ is $0$, which is exactly determined by $x_3$ of $\gamma_1$. 

\textit{Example 2.---} In the model-$[0,1,2,4]$ (to be studied in Sec.~\ref{subsec:fractonic}), the total space dimension $D=4$ and the leaf space dimension $d_l=2$. For a $0$-cube $\gamma_0=(0,0,0,0)$, there are $\binom{4-0}{2-0}=6$ leaves associated with it, which are respectively $\langle \hat{x}_1, \hat{x}_2\rangle$, $\langle  \hat{x}_1, \hat{x}_3\rangle$, $ \langle  \hat{x}_1, \hat{x}_4\rangle$, $\langle  \hat{x}_2, \hat{x}_3\rangle$, $ \langle  \hat{x}_2, \hat{x}_4\rangle$ and $\langle  \hat{x}_3, \hat{x}_4 \rangle$. Both   coordinate components $x_3$ and $x_4$ of each lattice site inside $\langle\hat{x}_1,\hat{x}_2\rangle$ are   $0$, which are exactly determined by $x_3,x_4$ of $\gamma_0$. 
	
\textit{Example 3.---} In the model-$[1,2,3,5]$ (to be studied in Sec.~\ref{subsec:fractonic}), the total space dimension $D=5$ and the leaf space dimension $d_l=3$. For a $2$-cube $\gamma_2=(0,0,0,\frac{1}{2},\frac{1}{2})$, there are $\binom{5-2}{3-2}=3$ leaves associated with it, which are respectively $ \langle  \hat{x}_1,\hat{x}_4,\hat{x}_5\rangle$, $ \langle  \hat{x}_2,\hat{x}_4,\hat{x}_5\rangle$ and $ \langle  \hat{x}_3,\hat{x}_4,\hat{x}_5\rangle$. Both   coordinate components $x_2$ and $x_3$ of each lattice site inside $\langle\hat{x}_1,\hat{x}_4,\hat{x}_5\rangle$ are   $0$, which are exactly determined by  $x_2,x_3$ of $\gamma_2$. 

Besides, as examples above suggest, the meaning of the $4$ indexes in the 4-tuple notation of models are listed below:
\begin{itemize}
	\item The first index $d_n$ is the dimension of the $d_n$-cube where a $B$-term in the hamiltonian is defined on.
	\item The second index $d_s$ is the dimension of the $d_s$-cube where a spin is defined on.
	\item The third index $d_l$ is the dimension of the leaf spaces.
	\item The fourth index $D$ is both the dimension of the $D$-cube where an $A$-term in the hamiltonian is defined on, and the dimension of the whole system.
\end{itemize}
A more detailed definition is given in Sec.~\ref{sec:family}.
 
\subsection{Definition of ``nearest'' via $L_1$-norm and $L_1$-distance}\label{sec_pre_distance}
$L_1$-distance \cite{boyd_vandenberghe_2004} is a distance function that is different from the Euclidean distance. In general, since there are various manners to define the length (a.k.a. norm) of a vector, and every well-defined length of the difference between two vectors can be used as a distance, we can use the $L_1$-norm to give the so-called $L_1$-distance. For vector $v=(x_1,x_2,...x_d)$, its $L_1$-norm is given by:
\begin{align}
L_1(v)=|x_1|+|x_2|+...|x_d|.
\end{align}
Then, by taking the $L_1$-norm of the difference between two vectors as a distance function, we obtain the $L^1$-distance between the two vectors. That is to say, for vector $v_1=(x_1,x_2,...x_d)$ and $v_2=(y_1,y_2,...y_d)$, their $L_1$-distance is given by:
\begin{align}
L_1(v_1,v_2)=|x_1-y_1|+|x_2-y_2|+...|x_d-y_d|.
\end{align}
In this article, we use $L_1$ distance to define whether two objects are ``nearest''. If an $m$-cube and an $n$-cube are said to be ``nearest'' to each other,  the $L_1$ distance should be: $L_1(\gamma_m,\gamma_n) = \frac{|m-n|}{2}$ when $m \neq n$. For $m=n$, we specially define two $n$-cubes being nearest if $L_1(\gamma_{n}^1,\gamma_{n}^2) = 1$.  

%

\subsection{Stacking of cubes: straight string, flat membrane, and beyond}

Moreover, in order to specify a region in hypercubic lattice, we also need a group of notations to denote   ``flat'' objects composed by $n$-cubes, like higher dimensional analogs of straight strings and flat membranes. Here, we define a $j$-dimensional analog of a straight string (in the original lattice) $S^j$ as a stack of nearest $j$-cubes  where all the $j$-cubes share the same coordinates along  orthogonal directions collected in the set    $\mathcal{C}^i_{\gamma_j}$. The simplest examples are straight lines $S^1$ and flat membranes $S^2$ as shown in Fig.~\ref{figure_cube}. In this figure, let us assume $D=3$. Then, $\mathcal{C}^i_{\gamma_1}=\{\hat{x}_2,\hat{x}_3\}$, $\mathcal{C}^i_{\gamma_2}=\{\hat{x}_3\}$. All $1$-cubes (i.e., links) in $S^1$ share same integer-valued coordinates  along both $\hat{x}_2$ and $\hat{x}_3$, and all $2$-cubes (i.e., plaquettes) in $S^2$ share same integer-valued coordinates  along $\hat{x}_3$/.   When we need to specify a $S^{i-1}$ which is located at the convergence of two $S^i$, we will also use $C^{i-1}$ to refer to it.

In a similar manner, we may define flat geometric objects in the dual lattice of the original lattice. More concretely, a $k$-dimensional analog (denoted by  $D^k$) of a flat membrane in the dual lattice can be defined as a stack of nearest $(D-k)$-cubes in the original lattice, where all the $(D-k)$-cubes share the same values for coordinates along orthogonal directions collected in the set  $\mathcal{C}^h_{\gamma_{D-k}}$. Alternatively speaking, a $D^k$ is just an $S^k$ if  the dual lattice and original lattice are switched. For instance, A $D^2$ in 3D space is a connected set of parallel links (i.e., $1$-cubes $\gamma_1$) all of which share the same $\mathcal{C}^h_{\gamma_1}$. The creation operator of fractons in the X-cube model is defined on a $D^2$ in 3D.

Specially, sometimes we may also use $D^k_p$ to denote a stack of nearest $p$-cubes in the original lattice, where all the $p$-cubes share the same values for coordinates along orthogonal directions collected in the set  $\mathcal{C}^h_{\gamma_{p}} \cup \mathcal{C}^{si}_{\gamma_{p}}$. Here $\mathcal{C}^{si}_{\gamma_{p}}$ is a subset of $\mathcal{C}^i_{\gamma_p}$ which satisfies $|\mathcal{C}^{si}_{\gamma_{p}}|=D-k-p$. Different from the previously defined objects, since $\mathcal{C}^{si}_{\gamma_p}$ is not completely specified, a $D^k_p$ can't be totally determined by $\gamma_p$, $k$ and $p$, so additional information is needed to specify a $D^k_p$. For example, in 3D X-cube model we will use $D^1_1$, i.e., $p=k=1$ and $D=3$. Let us consider a $1$-cube $\gamma_1=(\frac{1}{2},0,0)$, so the two sets of orthogonal directions are fixed: $\mathcal{C}^h_{\gamma_1}=\{\hat{x}_1\}$ and $\mathcal{C}^i_{\gamma_1}=\{\hat{x}_2,\hat{x}_3\}$. Therefore, $\mathcal{C}^{si}_{\gamma_1}=\{\hat{x}_2\}$ or $\{\hat{x}_3\}$, leading to two choices: $\mathcal{C}^h_{\gamma_{1}} \cup \mathcal{C}^{si}_{\gamma_{1}}=\{\hat{x}_1,\hat{x}_2\}$ or $\{\hat{x}_1,\hat{x}_3\}$. As a result,    there are two possible directions for stacking $1$-cubes in $D^1_1$, i.e., $\{(\frac{1}{2},0,j)|j=0,1,2,\cdots\}$ and $\{(\frac{1}{2},i,0)|i=0,1,2,\cdots\}$. When we need to specify a $D^k_p$ in the remainder of this article, additional information will always be given in the context.

As for boundaries,   the boundary $\partial S^1$ is simply given by the two endpoints of  $S^1$; the boundary    $\partial S^2$ is a closed string;  It's a bit   difficult to define the boundary of a $D^k$, but the vertices of $D^k$ can be naturally obtained by regarding   $D^k$ as a $k$-dimensional polytope.

\section{$[D-3,D-2,D-1,D]$ model series}
\label{sec:1234series}

\subsection{Construction of $[D-3,D-2,D-1,D]$ models}
\label{subsec:cons_1234}

In this section,  we first consider   models on a $D$-dimensional hypercubic lattice where    spins are located on $(D-2)$-cubes instead of links, while keeping the basic form of X-cube Hamiltonian unaltered. By introducing a 4-tuple notation, this consideration   is called $[D-3,D-2,D-1,D]$ models.  The Hamiltonian of general form is given by ($J>0$ and $K>0$ are always assumed):
\begin{align}
\label{eq:1234_branch_ham}
H_{D} = -J\sum_{\{{\gamma}_D\}} A_{{\gamma}_D} - K \sum_{\{{\gamma}_{D-3}\}} \sum_{l} B^l_{{\gamma}_{D-3}}\,,\end{align} 
where   $A_{{\gamma}_D}$ is the product of spin operators $\sigma^x$'s located on the centers of $(D-2)$ cubes that are  nearest\footnote{The accurate definition of ``nearest'' is given in Sec.~\ref{sec_pre_distance}} to hypercube ${\gamma}_D$.  In this series of models, a $B$ operator is associated with a $d_n$-cube and a leaf space $l$ with $d_n=D-3$ and $d_l=D-1$.   More concretly, $B^l_{{\gamma}_{D-3}}$ is the product of all $\sigma^z$'s which are not only nearest to ${\gamma}_{D-3}$ but also located inside  the leaf $l$.  The number of leaf spaces associated with each $\gamma_{d_n}$ is always $\binom{D-d_n}{d_l-d_n}\equiv\frac{(D-d_n)!}{(d_l-d_n)!(D-d_l)!}=\frac{3!}{2!}=3$ regardless of $D$.  In the following subsections, we will concentrate on $[0,1,2,3]$ and $[1,2,3,4]$ these two models to explore their excitation spectra. Especially there are simple dimension reduction rules for simple excitations in this model series as shown in Fig.~\ref{fig_tree}. The details are collected in Table~\ref{table:excitations}.
\begin{table}[htp] 	
	\caption{Typical examples of simple excitations (i.e., $\mathsf{E}^s$ sector) in $[D-3,D-2,D-1,D]$ models. }

	\begin{tabular}{cc c c}
		\hline
		
		\hline
	$D$    &~~~Excitations labels~~~& Creation operators \\
		\hline
		\multirow{2}{*}{3} &  $(0,0)$ & $\prod\limits_{{\gamma}_1 \in D^2} \sigma^z_{{\gamma}_1}$
		\\
		& $(0,1)$ & $\prod\limits_{{\gamma}_1 \in S^1} \sigma^x_{{\gamma}_1}$
		\\
		& $(0,2) $ & $\prod\limits_{{\gamma}_1 \in D^1_1}\sigma^z_{{\gamma}_1}$
		\\		 
		\hline
		\multirow{2}{*}{4} &  {$(0,0)$} & $\prod\limits_{{\gamma}_2 \in D^2} \sigma^z_{{\gamma}_2}$ 
		\\
		& $(1,2)$ & $\prod\limits_{{\gamma}_2 \in S^2} \sigma^x_{{\gamma}_2}$
		\\
		& $(0,3)   $ & $\prod\limits_{{\gamma}_2 \in D^1_2}\sigma^z_{{\gamma}_2}$
		\\		
		\hline
		\multirow{2}{*}{5} &  {$(0,0)$} & $\prod\limits_{{\gamma}_3 \in D^2} \sigma^z_{{\gamma}_3}$
		\\
		& $(2,3)$ & $\prod\limits_{{\gamma}_3 \in S^3} \sigma^x_{{\gamma}_3}$ 
		\\ 
		& $(0,4) $ & $\prod\limits_{{\gamma}_3 \in D^1_3}\sigma^z_{{\gamma}_3}$
		\\		
\hline

\hline	\end{tabular}
	\label{table:excitations}
\end{table}

\subsection{The X-cube model as $[0,1,2,3]$}
\label{subsec:rev_of_xcube}

\begin{figure}[t]
	\centering  
	\includegraphics[width=0.4\textwidth]{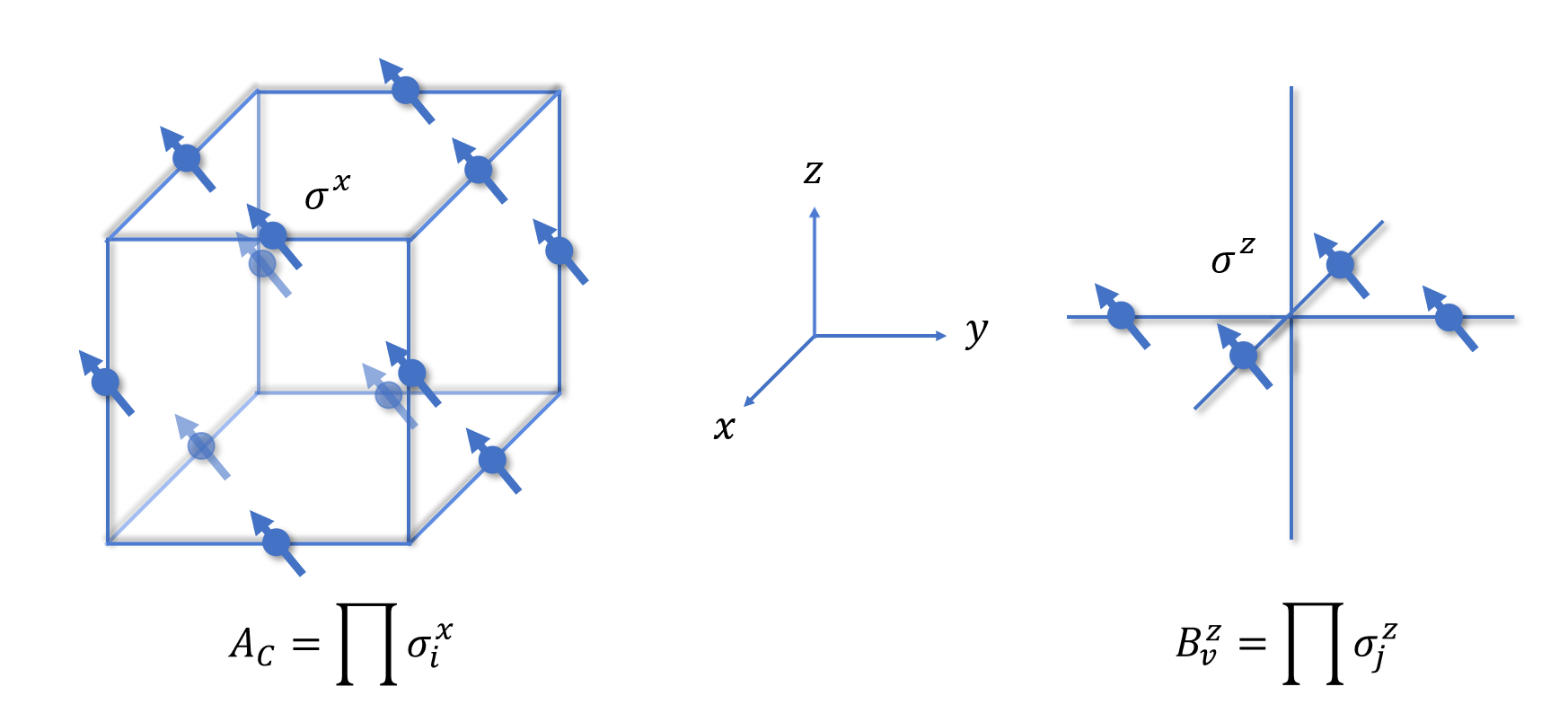}
	\caption{  Terms in the X-cube Hamiltonian. As shown in the figure, the $A_c$ term is composed of the 12 spins  ($\sigma^x$) on the edges of the cube, while the $B^z_v$ term is composed of the 4 spins ($\sigma^z$) on the legs of a vertex. For simplicity, only one of the 3 $B^i_v$ terms ($i=x,y,z$) on the vertex $v$ is shown, while $B_v^x=\prod \sigma^z$ and $B_v^y=\prod \sigma^z$ are not shown.}
	\label{fig_xcube}
\end{figure}

The well-understood X-cube model is labeled by $[0,1,2,3]$ in our notation. As the name suggests, X-cube model is defined on a cubic lattice, with $1/2$-spins sitting on  links. The Hamiltonian is of the form  \cite{Vijay2016}:
\begin{align}
\label{eq:-3-2-10Hamiltonian}
H_{X-cube} = -J\sum_{c} A_{c} - K \sum_{v} \sum_{i} B^i_v\, 
\end{align}
which is alternatively written as Eq.~(\ref{equivalent_form_xcube}) in terms of geometric notations. 
Here, the term $A_c$ of a given cube $c$ consists of the product of the $x$ components (i.e., $\sigma_x$) of the twelve spins around the cube $c$; $B^i_v$ means the product of $\sigma_z$'s  that are (i) inside the 2D plane that is perpendicular to the direction $i$ and (ii) nearest to the vertex $v$. The summation of $c$ and $v$ are respectively over all   cubes and vertices, while the summation of $i$ is over the three spatial dimensions.   The model is shown pictorially in Fig.~\ref{fig_xcube}.

  With the $\sigma^z$ basis, we can regard the links with $\sigma^z=-1$  as being ``occupied'' by strings and  the links with $\sigma^z=1$ spins as being ``unoccupied''. In this manner, the total Hilbert space can be alternatively represented by all kinds of different string configurations including both open and closed strings. Then, by solving the equations $A_c=1,\ \forall \ c $ and $B^i_v=1,\ \forall \ v$, with the open boundary condition, we can directly derive the ground state as $\ket{s_i}$: $\ket{\Phi}=\prod_{c} \frac{1+A_{c}}{\sqrt{2}} \ket{\uparrow \uparrow \uparrow\dots\uparrow},$ where $\ket{\uparrow \uparrow \uparrow\dots\uparrow}$ refers to the state with zero string. In the remainder of this article, $\ket{\uparrow \uparrow \uparrow\dots\uparrow}$ will be used as a reference state frequently.  The ground state of Eq.~(\ref{eq:-3-2-10Hamiltonian}) is dubbed as ``cage-net'' condensation  \cite{Prem2018}. If we consider the X-cube model on a 3-torus of the size $L\times L \times L$, the ground state will be degenerate, and the ground state degeneracy (GSD) is given by $\log_2 \text{GSD} = 6L-3$. The linear term here is also a significant feature of fracton orders, as it means that the GSD grows subextensively \cite{Vijay2016,Shirley2018}.

 \begin{table*}[htp] 	
	\caption{Typical examples of excitations in $[0,1,2,3]$ model. Simple excitations are labeled by a pair of integers. }
	
	\begin{tabular}{ccccc}
		\hline
		
		\hline
		Excitations & \begin{minipage}{1in} Sectors\end{minipage} &~~Flipped stabilizers~~& Creation operators \\
		\hline
		fracton:  {$(0,0)$} & $\mathsf{E}^s$ & $A_{\gamma_3}$ & $\prod\limits_{{\gamma}_1 \in D^2} \sigma^z_{{\gamma}_1}$ 
		\\
		lineon: $(0,1)$ & $\mathsf{E}^s$ & $B^l_{\gamma_0}$ & $\prod\limits_{{\gamma}_1 \in S^1} \sigma^x_{{\gamma}_1}$
		\\
		connected planeon: $(0,2)$& $\mathsf{E}^s$ & $A_{\gamma_3}$ & $\prod\limits_{{\gamma}_1 \in D^1_1}\sigma^z_{{\gamma}_1}$
		\\
		  
		disconnected planeon  & $\mathsf{E}^d$ & $A_{\gamma_3}$ & $\prod\limits_{{\gamma}_1 \in D^2} \sigma^z_{{\gamma}_1}$ 
		\\
		
		\hline
		
		\hline	
	\end{tabular}
	\label{table:0123_excitations}
\end{table*}

Some representative excitations of the X-cube model are summarized in Table \ref{table:0123_excitations}. In the X-cube model, there are two most important classes of   excitations---lineons and fractons, which are respectively originated from the eigenvalue flip of $B^i_v$ and $A_c$ terms.  Let us explain in details:
 \begin{itemize}
 \item \emph{An excited state with one lineon.} The $B^i_v=-1$ excitations, dubbed ``lineons'',  are generated by string operator $W(S^1)=\prod_{\gamma_{1} \in S^1} \sigma^x_{\gamma_{1}}$ composed of $\sigma^x_{\gamma_{1}}$ along the   open string $S^1$ which must be absolutely straight. The point-like excitations at the endpoints of a string are restricted in the line where the string sits, thus the name ``lineons''. In our notation, the end-of-string excitations are $(0,1)$-type point-like excitations and belong to $\mathsf{E}^s$. For example, if the string $S^1$ is along $\hat{x}$-axis, the eigenvalues of both $B^y_v$ and $B^z_v$ at each endpoint of $S^1$ will be flipped, rendering $2K$ energy cost.  Therefore,   there are in total three ``lineons'', denoted by $\ell_x$, $\ell_y$, and $\ell_z$, where the subscripts denote the directions of straight lines along which   lineons can move. 
 
 \item \emph{An excited state with two spatially separate lineons.} If $\ell_x$ and $\ell_y$ are able to meet at some point, they fuse into $\ell_z$ which is still a point, a zero-dimensional manifold. An excited state with \emph{these}  two spatially separate point-like pieces (i.e., $\ell_x$ and $\ell_y$)  belongs to $\widetilde{\mathsf{E}}^d$ which is, by definitions in Sec.~\ref{sec_intro_sector}, eventually $\mathsf{E}^s$.  But  $\ell_x$ and $\ell_y$ are unable to meet each other if the two straight lines do not intersect. If this is the case, the geometric shape of the excited state with a pair of $\ell_x$ and $\ell_y$  is intrinsically disconnected. As a result, such an  excited state belongs to $\mathsf{E}^d$ rather than $\widetilde{\mathsf{E}}^d$.  Likewise, if there are two $\ell_x$ lineons which move along two   parallel straight lines of $x$ direction, such an  excited state is also  in $\mathsf{E}^d$ since the geometric shape (two spatially separate points) are intrinsically disconnected.

\item \emph{An excited state with one fracton.} In addition to lineons, the $A_c=A_{\gamma_3}=-1$ excitations correspond to fractons (i.e. $(0,0)$-type excitations) associated to the cube $c$. Fractons, as point-like excitaitons, belong to $\mathsf{E}^s$. More precisely,  fractons are created by operators of the form $W(D^2)=\prod_{\gamma_1 \in D^2} \sigma^z_{\gamma_1}$, 
where $D^2$ is an absolutely flat 2-dimensional membrane in the dual lattice.  The cubes $c$'s are located at the corners of  $D^2$, each of which requires $J$ energy cost. For example, if $D^2$ is simply a rectangular, there will be four emerged fractons at the four corners. One can show that  fractons are totally immobile. More concretely,  moving a single fracton by applying spin operators will create additional new fractons nearby. 

\item \emph{An excited state with two fractons.} Despite that fractons are immobile, a pair of two nearby fractons generated by one membrane can move freely in the 2D plane perpendicular to the link between the two combined fractons. Thus these pairs, dubbed ``connected planeons'', are identified as $(0,2)$-type excitations in $\mathsf{E}^s$ sector in our notation when the component fractons are exactly next to each other.  If the two component fractons are separate, the corresponding excitation is called ``disconnected planeons'' which belong to $\mathsf{E}^d$. Two types of planeons cannot be changed to each other by local operators since they belong to different sectors of Hilbert space.
\end{itemize}
   
%
%
%
%
%
%
%

  \subsection{Simple excitations in the model-$[1,2,3,4]$}
\label{subsec:ex_1234}

\begin{figure*}[t]
	\subfigure[A ${\gamma}_{D-3}$ in ${[0,1,2,3]}$ model]{
		\label{fig_graph_p}
		\includegraphics[width=0.22\textwidth]{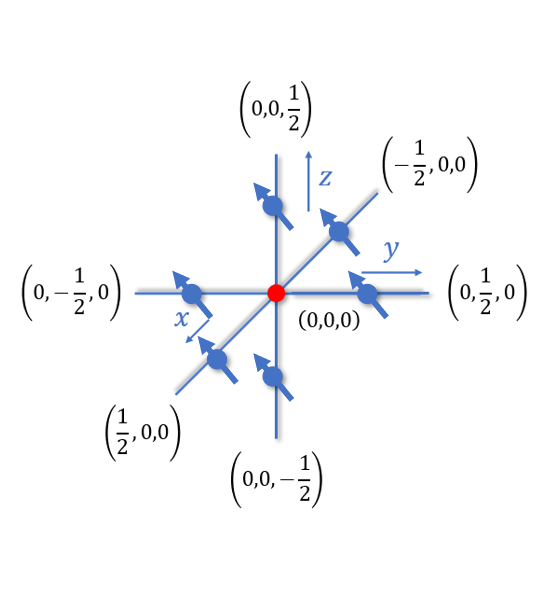}}
	\subfigure[Graphical representation of a ${\gamma}_{D-3}$ in ${[0,1,2,3]}$ model]{
		\label{fig_graph_3d}
		\includegraphics[width=0.22\textwidth]{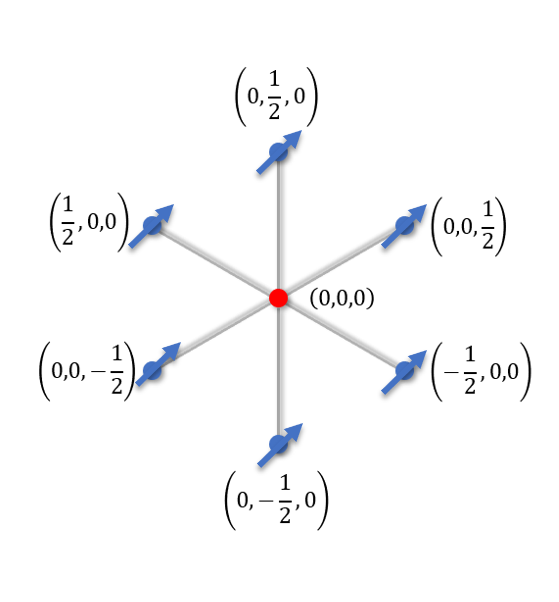}}
	\subfigure[Graphical representation of a ${\gamma}_{D-3}$ in ${[1,2,3,4]}$ model]{
		\label{fig_graph_4d}
		\includegraphics[width=0.22\textwidth]{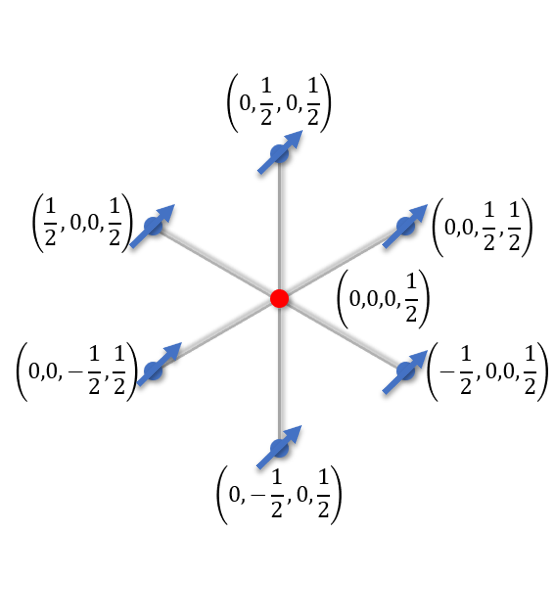}}
	\subfigure[Graphical representation of a ${\gamma}_{D-3}$ in ${[2,3,4,5]}$ model]{
		\label{fig_graph_5d}
		\includegraphics[width=0.22\textwidth]{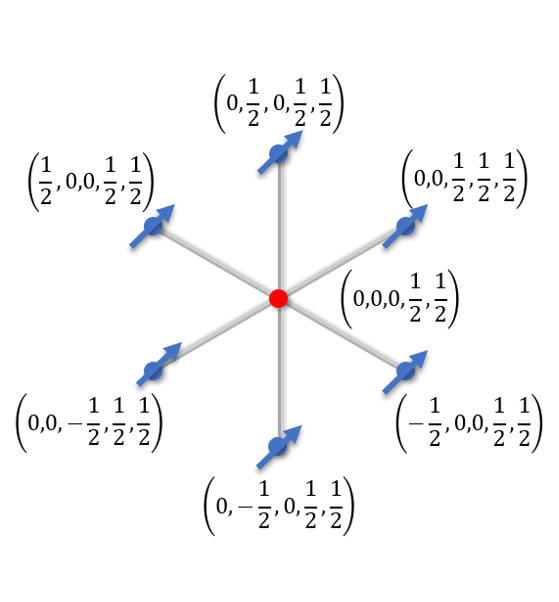}}
	
	\caption{   Graphical representation of ${\gamma}_{D-3}$ in some $[D-3,D-2,D-1,D]$ models. Here, (b) is the graph representation of the ${\gamma}_{D-3}=(0,0,0)$ presented in (a). As we can see, in higher dimensions, the graph representation of a ${\gamma}_{D-3}$ is completely the same as in the 3D case: a ${\gamma}_{D-3}$ always connects 3 pairs of $\gamma_{D-2}$, which means 3 pairs of spins in $[D-3,D-2,D-1,D]$ models.}
	\label{fig_graph}
\end{figure*}

In the remainder of this section, we focus on the model-$[1,2,3,4]$. In this model,  for a specific $4$-cube    $\gamma_4=(\frac{1}{2},\frac{1}{2},\frac{1}{2},\frac{1}{2})$ and  a specific $1$-cube $\gamma_1=(0,0,0,\frac{1}{2})$ respectively, we have\footnote{See the ``Example 1'' in Sec.~\ref{sec_pre_leaf} for an introduction to the leaves in model-$[1,2,3,4]$.}
\begin{align}
	\label{eq:A terms}
	\begin{split}
	A_{(\frac{1}{2},\frac{1}{2},\frac{1}{2},\frac{1}{2})} =& \sigma^x_{(0,0,\frac{1}{2},\frac{1}{2})} \sigma^x_{(0,1,\frac{1}{2},\frac{1}{2})}  \sigma^x_{(1,0,\frac{1}{2},\frac{1}{2})}  \sigma^x_{(1,1,\frac{1}{2},\frac{1}{2})}\\ &\sigma^x_{(0,\frac{1}{2},0,\frac{1}{2})}  \sigma^x_{(0,\frac{1}{2},1,\frac{1}{2})} 
	\sigma^x_{(1,\frac{1}{2},0,\frac{1}{2})} \sigma^x_{(1,\frac{1}{2},1,\frac{1}{2})}\\    &\sigma^x_{(0,\frac{1}{2},\frac{1}{2},0)} \sigma^x_{(0,\frac{1}{2},\frac{1}{2},1)}  \sigma^x_{(1,\frac{1}{2},\frac{1}{2},0)} \sigma^x_{(1,\frac{1}{2},\frac{1}{2},1)}\\  &\sigma^x_{(\frac{1}{2},0,0,\frac{1}{2})}  \sigma^x_{(\frac{1}{2},0,1,\frac{1}{2})}  \sigma^x_{(\frac{1}{2},1,0,\frac{1}{2})} \sigma^x_{(\frac{1}{2},1,1,\frac{1}{2})}\\  &\sigma^x_{(\frac{1}{2},0,\frac{1}{2},0)} \sigma^x_{(\frac{1}{2},0,\frac{1}{2},1)} \sigma^x_{(\frac{1}{2},1,\frac{1}{2},0)}  \sigma^x_{(\frac{1}{2},1,\frac{1}{2},1)}\\  &\sigma^x_{(\frac{1}{2},\frac{1}{2},0,0)} \sigma^x_{(\frac{1}{2},\frac{1}{2},0,1)}  \sigma^x_{(\frac{1}{2},\frac{1}{2},1,0)} \sigma^x_{(\frac{1}{2},\frac{1}{2},1,1)},\\
	\end{split}
\end{align} 
and 
\begin{align}
	\label{eq:B terms}
	\begin{split}
	&B^{\langle \hat{x}_1,\hat{x}_2,\hat{x}_4\rangle}_{(0,0,0,\frac{1}{2})} = \sigma^z_{(\frac{1}{2},0,0,\frac{1}{2})} \sigma^z_{(-\frac{1}{2},0,0,\frac{1}{2})} \sigma^z_{(0,\frac{1}{2},0,\frac{1}{2})} \sigma^z_{(0,-\frac{1}{2},0,\frac{1}{2})},\\
	&B^{\langle \hat{x}_2, \hat{x}_3, \hat{x}_4\rangle}_{(0,0,0,\frac{1}{2})} = \sigma^z_{(0,\frac{1}{2},0,\frac{1}{2})} \sigma^z_{(0,-\frac{1}{2},0,\frac{1}{2})} \sigma^z_{(0,0,\frac{1}{2},\frac{1}{2})} \sigma^z_{(0,0,-\frac{1}{2},\frac{1}{2})},\\
	&B^{\langle \hat{x}_1, \hat{x}_3, \hat{x}_4\rangle}_{(0,0,0,\frac{1}{2})} = \sigma^z_{(\frac{1}{2},0,0,\frac{1}{2})} \sigma^z_{(-\frac{1}{2},0,0,\frac{1}{2})} \sigma^z_{(0,0,\frac{1}{2},\frac{1}{2})} \sigma^z_{(0,0,-\frac{1}{2},\frac{1}{2})}.\\
	\end{split}
\end{align} 
Although the model looks strange at first sight, it is just a generalization of the 3D X-cube model given by Eq.~(\ref{eq:-3-2-10Hamiltonian}) and its equivalent form Eq.~(\ref{equivalent_form_xcube}) (another 4D generalization is the the model-$[0,1,2,4]$, which is discussed in Sec.~\ref{subsec:fractonic}). As we can see that once we choose $D=3$ in Eq.~(\ref{eq:1234_branch_ham}), the model would simply reduce to Eq.~(\ref{equivalent_form_xcube}). In other words, the X-cube model is the simplest case in the series $[D-3,D-2,D-1,D]$. Furthermore,  an $A_{{\gamma}_D}$ always overlaps with a nearest $B$ operator by even number of spins, as an $A_{{\gamma}_D}$ always covers one of each pair of spins linked by a nearest ${\gamma}_{D-3}$, and a $B$ operator is composed of 2 such pairs. Therefore, our generalized models are still exactly solvable. Fig.~\ref{fig_graph} gives a graph demonstration.

The ground state configuration  must satisfy the following conditions: $A_{{\gamma}_D} \ket{\phi} = \ket{\phi},\ B^l_{{\gamma}_{D-3}} \ket{\phi} = \ket{\phi}\,, \forall\ {\gamma}_D, \ {\gamma}_{D-3},\ l,$. Topological excitations   appear in the region where one or a proper combination of these conditions is violated.   In the $\sigma^z$ basis, we can regard the ground states as condensations of ``$D$-cage nets'', where ``$D$-cage'' is the $D$-dimensional analog of the ``cage'' proposed in Ref.~\cite{Prem2018}. 
 When the boundary of the system is open, we can obtain the ground state wave function as the equal-weight superposition of all $D$-cages: $\ket{\Phi}=\prod_{\gamma_D} \frac{1+A_{\gamma_D}}{\sqrt{2}} \ket{\uparrow \uparrow \uparrow\dots\uparrow}$. Here $\ket{\uparrow \uparrow \uparrow\dots\uparrow}$ is a reference state where spins are all upward along $z$-axis.

Next, we move on to the excitation spectrum of the model Hamiltonian given by Eq.~(\ref{eq:1234_branch_ham}). We shall begin with the energy cost of ``simple excitations''. When the lattice constant $a$ goes to $0$, these excitations will look like some    connected manifolds, like points, strings, membranes and so on. Elementary introductions to manifold can be found on Page 219 of Ref.~\cite{EGUCHI1980213}.

Analogous to the original X-cube model, the most representative simple excitations in the model-$[1,2,3,4]$ can be classified into two classes: $(0,0)$-type  excitations  and $(1,2)$-type  excitations. The former are excited by operators $W(D^2)=\prod_{{\gamma_2}\in D^2} \sigma^z_{\gamma_2}$, resulting in eigenvalue flip (i.e.,  $1\longrightarrow -1$) of $A_{\gamma_4}$  for $\gamma_4$'s at the corners of the $D^2$. The latter are excited by $W(S^2)=\prod_{{\gamma_2}\in S^2} \sigma^x_{\gamma_2}$, resulting in eigenvalue flip (i.e., $1\longrightarrow -1$) of $B^l_{\gamma_1}$ for $\gamma_1$'s along $\partial S^2$.  For the sake of convenience, we will use the expressions $A=-1$ and $B=-1$ to describe such eigenvalue flip. The general definition of the notations $D^2$ and $S^2$ can be found in Sec.~\ref{sec:rev_not}.   Starting from the next subsection, we will discuss the excitation spectrum of this model systematically. Some excitations are collected in Table~\ref{table:1234_excitations}.
\begin{table*}[htp] 	
	\caption{ Typical examples of  excitations in the model-$[1,2,3,4]$. Note: excitations  in Sec.~\ref{sec_multiple_loop} are collected separately in Table~\ref{table:comparison_of_fusion_results}.}
	
	\begin{tabular}{ccccc}
		\hline
		
		\hline
		Excitations &~~~Sectors~~~& Flipped stabilizers & Creation operators \\
		\hline
		 fracton: {$(0,0)$} & $\mathsf{E}^s$ & $A_{\gamma_4}$ & $\prod\limits_{{\gamma}_2 \in D^2} \sigma^z_{{\gamma}_2}$ 
		\\
		$(1,2)$ & $\mathsf{E}^s$ & $B^l_{\gamma_1}$ & $\prod\limits_{{\gamma}_2 \in S^2} \sigma^x_{{\gamma}_2}$
		\\
\begin{minipage}{1.5in}connected  volumeon: $(0,3)$ \end{minipage}& $\mathsf{E}^s$ & $A_{\gamma_4}$ & $\prod\limits_{{\gamma}_2 \in D^1_2}\sigma^z_{{\gamma}_2}$
		\\
\\		chairon & $\mathsf{E}^c$ & $B^l_{\gamma_1}$ & \begin{minipage}{2in}$\prod_{{\gamma_2}\in S^2_I} \sigma^x_{\gamma_2} \prod_{{\gamma_2}\in S^2_{II}} \sigma^x_{\gamma_2}$, \\where $\partial S^2_{I} \cap \partial S^2_{II} = C^1 \neq \emptyset$
		\end{minipage}\\ ~\\
		yuon & $\mathsf{E}^c$ & $B^l_{\gamma_1}$ &  \begin{minipage}{2in} $\prod_{{\gamma_2}\in S^2_I} \sigma^x_{\gamma_2} \prod_{{\gamma_2}\in S^2_{II}} \sigma^x_{\gamma_2} \prod_{{\gamma_2}\in S^2_{III}} \sigma^x_{\gamma_2}$, where $\partial S^2_{I} \cap \partial S^2_{II} \cap \partial S^2_{III} = C^1 \neq \emptyset$
		\end{minipage}\\
		\\
		disconnected volumeon & $\mathsf{E}^d$ & $A_{\gamma_4}$ & $\prod\limits_{{\gamma}_2 \in D^2} \sigma^z_{{\gamma}_2}$ 
		\\

		\hline
		
		\hline	
	\end{tabular}
	\label{table:1234_excitations}
\end{table*}

%
%
%

\subsubsection{$(0,0)$-type point-like excitations (fractons)}

 Firstly, let's consider the $(0,0)$-type excitations, i.e., fractons. When we act $W(D^2)=\prod_{{\gamma_2}\in D^2} \sigma^z_{\gamma_2}$ on the ground state,  the minimal polytope $P$ that envelops all the spins ($\sigma^z$'s) acted on by $W(D^2)$ is   4-dimensional. Obviously, all   $4$-cube operators $A_{{\gamma}_4}$ inside $P$ will contain even number of  $\sigma^x$'s that are acted on by $W(D^2)$, which keeps eigenvalues of all such operators $A_{\gamma_4}$ unaltered, i.e., $A_{\gamma_4}=1$ for $\gamma_4\in P$. Nevertheless,    for all $\gamma_4$'s that sit on the corners (i.e., vertices of $D^2$)  have only one  spin  per ${\gamma_4}$ that is acted on by $W(D^2)$, which flips the eigenvalue of these $A_{\gamma_4}$, i.e., $A_{\gamma_4}=-1$ for such $\gamma_4$'s. 
 
\begin{figure}[t]
	\centering  
	\subfigure[Fractons in ${[1,2,3,4]}$ model]{
		\label{fig_2a}
		\includegraphics[width=0.28\textwidth]{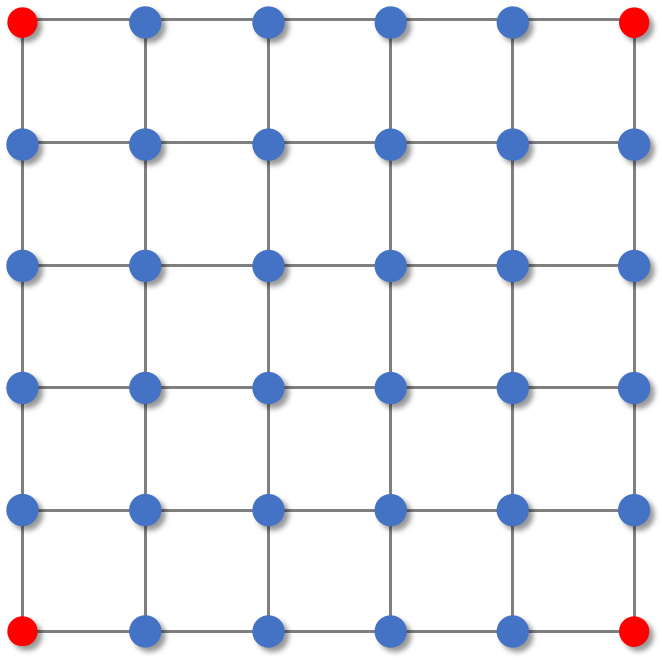}}
	\subfigure[A closed string excitation in ${[1,2,3,4]}$ model]{
		\label{fig_2b}
		\includegraphics[width=0.32\textwidth]{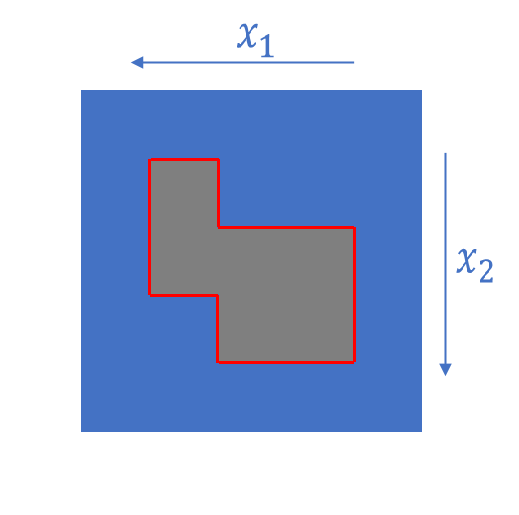}}
	\caption{  Fractons and string excitations in the model ${[1,2,3,4]}$. (a) demonstrates the distribution of fractons excited by $W(D^2)=\prod_{{\gamma_2}\in D^2} \sigma^z_{\gamma_2}$. Flipped spins sitting on $2$-cubes represented by dots are distributed on the sites of a $2$-dimensional array, which is exactly the $D^2$. The red dots at the corners of the $D^2$ refer to the spins that are nearest to an excited nearest $4$-cube operator $A_{{\gamma}_d}$, and the blue dots refer to the spins  whose nearest $A_{{\gamma}_d}$'s are not excited at all. In (b) the blue plaquettes label untouched spins while the grey plaquettes label the spins on which $W(D^2)$ acts. The  string excitation  as the domain wall  is highlighted with red lines.}
	\label{fig_2}
\end{figure}

 For example, we can apply $W(D^2)=\prod_{\gamma_1 \in D^2} \sigma^z_{\gamma_1}$ on the ground state, where, according to the definition in Sec.~\ref{sec:rev_not}, $D^2=\{(i, j, \frac{1}{2}, \frac{1}{2})|i,j=0,1,\cdots,L\}$.  Geometrically, $D^2$ forms  a square of $L\times L$, $L\in\Z$. For any hypercube $(m+\frac{1}{2}, n+\frac{1}{2}, \frac{1}{2}, \frac{1}{2})$, where $m,n=0,1,2,\dots,L-1$, there are always four spins located at respectively $(m, n, \frac{1}{2}, \frac{1}{2})$, $(m+1, n, \frac{1}{2}, \frac{1}{2})$,$(m, n+1, \frac{1}{2}, \frac{1}{2})$ and $(m+1, n+1, \frac{1}{2}, \frac{1}{2})$ that are acted on by $W(D^2)$. Therefore, the associated operators $A_{\gamma_4}$ have their eigenvalues unchanged , i.e., $A_{\gamma_4}=1$. Only for the $4$-cubes at the corners, like  $\gamma_4=(-\frac{1}{2},-\frac{1}{2},\frac{1}{2},\frac{1}{2})$, there is  just one spin per $\gamma_4$ acted on by $W(D^2)$, thus $A_{\gamma_4}=-1$ (i.e., we can say the operator is excited). As a result, it's straightforward to conclude that these excitations are of $(0,0)$-type, as any movement of such an excitation will  create more corners associated with  additional excitations and energy cost. See Fig.~\ref{fig_2a} for a schematic demonstration. 

\subsubsection{$(0,3)$-type point-like excitations (connected volumeons)}

In the model ${[1,2,3,4]}$, a pair of two fractons at two neighbouring corners of a membrane in the dual lattice are not a planeon anymore. Instead, these pairs become ``volumeons'' for an observer   in a 4D world. While it should be noticed that such a pair in general can belong to either $\mathsf{E}^s$ sector or $\mathsf{E}^d$ sector, and here we only consider the $\mathsf{E}^s$-type pair in which two $\gamma_4$ associated with the two fractons are nearest to each other, so we can label it by $(0,3)$. This excited state is called ``connected volumeons'', analogous to ``connected planeons'' in Table~\ref{table:0123_excitations}.  For instance, we can consider acting the open string operator  $W(D^1_2)=\prod_{\gamma_2 \in D^1_2} \sigma^z_{\gamma_2}$ on the ground state, where $D^1_2=\{(0,i,\frac{1}{2},\frac{1}{2})|i=0,1,2,3,\dots L-1,L\}$.\footnote{Here ``string'' means that all spins acted on by the operator form an open string. The precise definition of $D^1_2$ is given in Sec.~\ref{sec:rev_not}). }  After that, in the neighborhood of an endpoint of the $D^1_2$, e.g.,  $(0,0,\frac{1}{2},\frac{1}{2})$, there are two nearest $4$-cube operators $A_{\gamma_4}$ with   $\gamma_4=(-\frac{1}{2},-\frac{1}{2},\frac{1}{2},\frac{1}{2})$ and $\gamma_4=(\frac{1}{2},-\frac{1}{2},\frac{1}{2},\frac{1}{2})$ whose eigenvalues are flipped. These two $4$-cubes form  a pair of nearest fractons, whose energy is $2J$.  Define a vector $\mathbf{r}$ connecting the two $4$-cubes: $\mathbf{r}=(\frac{1}{2}-(-\frac{1}{2}),-\frac{1}{2}-(\frac{1}{2}),\frac{1}{2}-\frac{1}{2},\frac{1}{2}-\frac{1}{2})=(1,0,0,0)$. Then, the pair of fractons can be  regarded as a dipole whose moment point in the direction $\mathbf{r}$, i.e., $\hat{x}_1$.

For the issue of mobility, let us attempt   to act $\sigma^z_{(0,-\frac{1}{2},0,\frac{1}{2})}$ to move the pair out of the line where the string is located at. As we can see, since $\sigma^z_{(0,-\frac{1}{2},0,\frac{1}{2})}$ flips the sign of $A_{\gamma_4}$ for $\gamma_4 \in \{(\frac{1}{2},-\frac{1}{2},\frac{1}{2},\frac{1}{2})$, $(-\frac{1}{2},-\frac{1}{2},-\frac{1}{2},\frac{1}{2})$, $(\frac{1}{2},-\frac{1}{2},-\frac{1}{2},\frac{1}{2})$, $(-\frac{1}{2},-\frac{1}{2},\frac{1}{2},\frac{1}{2})\}$, $\sigma^z_{(0,-\frac{1}{2},0,\frac{1}{2})}$ can move the pair along $\hat{x}_4$ direction. Since $(0,-\frac{1}{2},0,\frac{1}{2})$ and $(0,-\frac{1}{2},\frac{1}{2},0)$ are symmetric about the string, the pair  can also be moved along the $\hat{x}_3$ direction.  As a result, the mobility of the  pair is restricted in the $3$-dimensional leaf space $\langle \hat{x}_2, \hat{x}_3, \hat{x}_4 \rangle$ with $x_1=0$.  
\subsubsection{$(1,2)$-type  string excitations of $6$ flavors}

\begin{figure*}[t]
	\centering  
	\subfigure[A string excitation labeled by $(1,2)^{\hat{x}_1,\hat{x}_2}$]{
		\label{fig_3a}
		\includegraphics[width=0.28\textwidth]{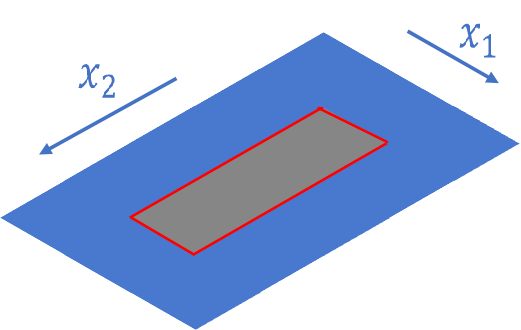}}
	\subfigure[Folding $(1,2)^{\hat{x}_1,\hat{x}_2}$ at the price of additional energy cost along the yellow crease.]{
		\label{fig_3b}
		\includegraphics[width=0.28\textwidth]{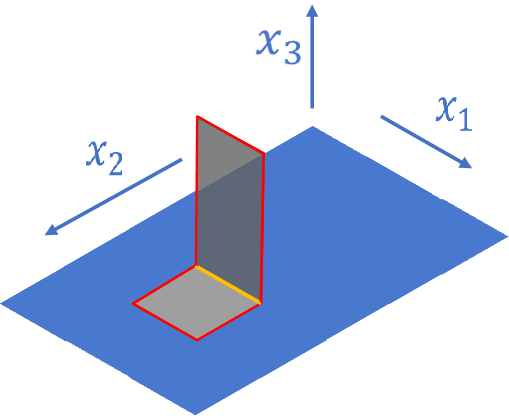}}
	\subfigure[A folded loop in pure topological orders]{
		\label{fig_3c}
		\includegraphics[width=0.28\textwidth]{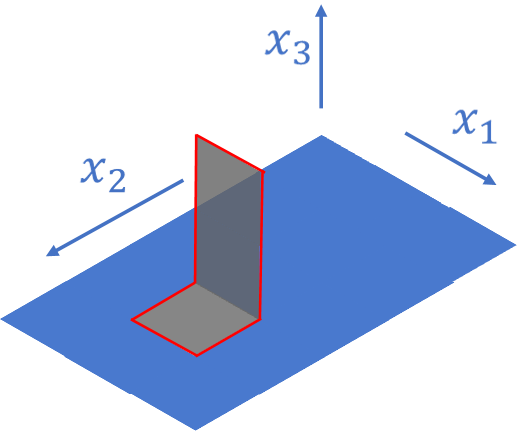}}

\caption{A $(1,2)^{\hat{x}_1,\hat{x}_2}$-type excitation cannot freely escape from the plane $\hat{x}_1$-$\hat{x}_2$. (a) A string excitation (marked by the red line) with restricted mobility and deformability, labeled by $(1,2)^{\hat{x}_1,\hat{x}_2}$.  By applying proper local operators, a part of  the string moves to $\hat{x}_1$-$\hat{x}_3$ plane but at the unavoidable localized energy price along the crease  line in yellow in (b), which leads to the deformability restriction.  Alternatively speaking,   the state in (b) is no longer a  single-string excitation. It is similar to moving a single fracton, where the consequence is to arrive at a state with multiple fractons. 
For comparison, if  (a) is produced in   pure topological order,   (a) can be deformed to (c)   such that the shape of a loop is gradually and freely deformed out of the plane.}
	\label{fig_3}
\end{figure*}

 Next we consider   excitations associated with flipped eigenvalues of $B^l_{\gamma_1}$. For each $\gamma_1$, there are three associated leaves labeled by $l$. We find that there are 6 flavors of $(1,2)$-type excitations--- string excitations\footnote{Both words ``string'' and ``loop'' will be used for the name of such excitations.}  that are created and moved within a certain plane \textit{only}, i.e.,  $\hat{x}_1$-$\hat{x}_2$, $\hat{x}_1$-$\hat{x}_3$, $\hat{x}_1$-$\hat{x}_4$, $\hat{x}_2$-$\hat{x}_3$, $\hat{x}_2$-$\hat{x}_4$, and $\hat{x}_3$-$\hat{x}_4$. An example is given in Fig.~\ref{fig_2b} (see also Fig.~\ref{fig_3a}). We use the symbol ``$  (1,2)^{\hat{x}_i,\hat{x}_j}$'' with two integers $1\leq i<j\leq 4$ to specify flavors. 
 
 More concretely, let us apply  an open membrane operator $W(S^2)=\prod_{{\gamma_2}\in S^2} \sigma^x_{\gamma_2}$ on the ground state.  For $\gamma_1$'s at the interior of $S^2$, by noting that  there    always exists exactly one pair of spins linked by each ${\gamma}_{1}$ being acted on by $W(S^2)$, the associated operators $B^l_{\gamma_1}$  will keep their eigenvalues (i.e., $B=1$) unaltered after $W(S^2)$ is applied.  Only for   ${\gamma}_{1} \in \partial S^2$, i.e., $\gamma_1$'s that form the boundary of $S^2$, the eigenvalues of the associated operators $B^l_{{\gamma}_{1}}$ will be flipped, i.e., $B=-1$, as shown in  Fig.~\ref{fig_2b}. That is to say, these   $B^l_{{\gamma}_{1}}$ operators with flipped eigenvalues constitute string excitations, of which the energy cost (i.e., excitation energy) is proportional to the length of the string. 
 
 Analogous to X-cube model,  $W(S^2)$ here can be classified into $6$ ``flavors'' according to $6$ different planes (i.e., $\hat{x}_1$-$\hat{x}_2$, $\hat{x}_1$-$\hat{x}_3$, $\hat{x}_1$-$\hat{x}_4$, $\hat{x}_2$-$\hat{x}_3$, $\hat{x}_2$-$\hat{x}_4$, and $\hat{x}_3$-$\hat{x}_4$) where  $S^2$ is located, and $W(S^2)$ of different flavors will flip different combinations of $B^l_{\gamma_1}$'s. In general, after applying $W(S^2)$ with $S^2$ being inside the $\hat{x}_i$-$\hat{x}_j$ plane, there will be exactly two flipped $B^l_{\gamma_1}$ terms at each $\gamma_1$ along $\partial S^2$, i.e.,    $B^{\langle \hat{x}_i, \hat{x}_j,\hat{x}_k \rangle}_{\gamma_1}$ and $B^{\langle \hat{x}_i, \hat{x}_j,\hat{x}_h \rangle}_{\gamma_1}$. Here $i,j,k,h \in \{1,2,3,4\}$ and $i,j,k,h$ are all different from each other. For example, by acting $W(S^2)$ on the ground state, where $S^2=\{(n+\frac{1}{2}, m+\frac{1}{2}, 0, 0)|m,n=0,1,2,\dots, L-1\}$, for an arbitrary $\gamma_1$ along the boundary of $S^2$, the eigenvalues of both $B^{\langle \hat{x}_1, \hat{x}_2,\hat{x}_3 \rangle}_{\gamma_1}$ and $B^{\langle \hat{x}_1, \hat{x}_2,\hat{x}_4 \rangle}_{\gamma_1}$ will be flipped. As a result, we find that the energy cost of this string excitation labeled by $(1,2)^{\hat{x}_1,\hat{x}_2}$ in the model-$[1,2,3,4]$ is $2KL$, where $L$ is the length of the string.  Before moving forward, let us summarize the ``stabilizers'' whose eigenvalues are flipped for $(1,2)$-type excitations of each flavor ($\gamma_1\in\partial S^2$):  
 \begin{itemize}
 \item $\mathbf{(1,2)^{\hat{x}_1,\hat{x}_2}}$: $B^{\langle\hat{x}_1,\hat{x}_2,\hat{x}_3\rangle}_{\gamma_1}$ and $B^{\langle\hat{x}_1,\hat{x}_2,\hat{x}_4\rangle}_{\gamma_1}$ (note: an example is given in Fig.~\ref{fig_3a}.)
 \item $\mathbf{(1,2)^{\hat{x}_1,\hat{x}_3}}$: $B^{\langle\hat{x}_1,\hat{x}_2,\hat{x}_3\rangle}_{\gamma_1}$ and  $B^{\langle\hat{x}_1,\hat{x}_3,\hat{x}_4\rangle}_{\gamma_1}$
 \item $\mathbf{(1,2)^{\hat{x}_1,\hat{x}_4}}$: $B^{\langle\hat{x}_1,\hat{x}_2,\hat{x}_4\rangle}_{\gamma_1}$ and  $B^{\langle\hat{x}_1,\hat{x}_3,\hat{x}_4\rangle}_{\gamma_1}$
 \item $\mathbf{(1,2)^{\hat{x}_2,\hat{x}_3}}$: $B^{\langle\hat{x}_1,\hat{x}_2,\hat{x}_3\rangle}_{\gamma_1}$ and  $B^{\langle\hat{x}_2,\hat{x}_3,\hat{x}_4\rangle}_{\gamma_1}$
 \item $\mathbf{(1,2)^{\hat{x}_2,\hat{x}_4}}$: $B^{\langle\hat{x}_1,\hat{x}_2,\hat{x}_4\rangle}_{\gamma_1}$ and  $B^{\langle\hat{x}_2,\hat{x}_3,\hat{x}_4\rangle}_{\gamma_1}$
 \item $\mathbf{(1,2)^{\hat{x}_3,\hat{x}_4}}$: $B^{\langle\hat{x}_1,\hat{x}_3,\hat{x}_4\rangle}_{\gamma_1}$ and  $B^{\langle\hat{x}_2,\hat{x}_3,\hat{x}_4\rangle}_{\gamma_1}$\,
 \end{itemize}

For the issue of mobility and deformability, the string excitation has a novel property here: it is restricted in the 2D plane where $S^2$ lies. Without loss of generality, as shown in Fig.~\ref{fig_3}, let us try to move the string excitation out of the plane where $S^2$ lies, by folding $S^2$  in $\hat{x}_1-\hat{x}_2$ plane  into $S^2_I$ in $\hat{x}_1-\hat{x}_3$ plane and $S^2_{II}$ in    $\hat{x}_1-\hat{x}_2$ plane. The   crease line denoted by $C^1$ is along $\hat{x}_1$ direction.  Nevertheless,  this process will cost additional energy  localized along $C^1$. Therefore, moving or deforming the $(1,2)$-type excitation out of the original plane is forbidden. Alternatively speaking, folding the loop in (a) only results in the state in (b) which is   not a single loop state, thus deformation out of plane is forbidden. So, what does free deformation look like? In a pure topological order, a loop in (a) can be sent to (c) by local operators such that the loop gradually evolves into $\hat{x}_1$-$\hat{x}_2$ plane without any obstruction.  In Sec.~\ref{subsec:com_ex_1234}, we will see that  the excited state represented by Fig.~\ref{fig_3b} actually belongs to $\mathsf{E}^c$ sector.  
 
But can the string excitation move and deform freely within the plane where it is located? It is easy to see that by applying $W(S^2)$ one can change the geometric  shape of the string within the same plane. Moreover,  no additional energy cost is required as long as the total length $L$ is unchanged.  In this sense, the string excitation can move and deform freely within a 2D subspace, so that such string excitations in our notation are labeled by $(1,2)$. For instance, let us  apply 
\begin{equation}
\label{eq_S1}
W(S^2_{I})=\prod_{{\gamma_2}\in S^2_{I}} \sigma^x_{\gamma_2} 
\end{equation}
on the ground state, where $S^2_{I}=\{(n+\frac{1}{2}, m+\frac{1}{2}, 0, 0)|m,n=0,1,2,\dots, L-1\}$. For an arbitrary $1$-cube, say $(\frac{3}{2},1,0,0)$ inside $S^2_{I}$, we can easily check that $W(S^2_{I})$ acts on two nearest spins at $(\frac{3}{2},\frac{1}{2},0,0)$ and $(\frac{3}{2},\frac{3}{2},0,0)$, so there will be no excited $B^l_{(\frac{3}{2},1,0,0)}$. While for $\gamma_1=(\frac{1}{2},0,0,0)$ on the boundary of $S^2_I$, since $W(S^2_{I})$ only acts on one nearest spin (at  $(\frac{1}{2},\frac{1}{2},0,0)$), so two $B^l_{(\frac{1}{2},0,0,0)}$ terms will be excited. Immediately after applying $W(S^2_{I})$ on the ground state, we   apply 
\begin{equation}
\label{eq_S2}
W(S^2_{II})=\prod_{{\gamma_2}\in S^2_{II}} \sigma^x_{\gamma_2}\,,
\end{equation}
where $S^2_{II}=\{(h+\frac{1}{2},0, k+\frac{1}{2}, 0)|h,k=1,2,\dots,L-1\}$. For $\gamma_1=(\frac{1}{2},0,0,0) \in \partial S^2_{I} \cap \partial S^2_{II}$, there will be still two excited $B^l_{(\frac{1}{2},0,0,0)}$ terms, which are respectively $B^{\langle \hat{x}_1 ,\hat{x}_2 ,\hat{x}_4 \rangle}_{(\frac{1}{2},0,0,0)}$ and $B^{\langle \hat{x}_1 ,\hat{x}_3 ,\hat{x}_4 \rangle}_{(\frac{1}{2},0,0,0)}$.

\subsection{Complex excitations in the model-$[1,2,3,4]$}
\label{subsec:com_ex_1234}
\subsubsection{Chairons of 12 flavors}

In the above discussions, we have analyzed   three types of simple excitations in the model-$[1,2,3,4]$: fractons $(0,0)$, volumeons $(0,3)$, and strings $(1,2)$.  All these excitations belong to the category of ``simple excitations'' ($\mathsf{E}^s$) as they are just simple geometric objects like points and strings. Surprisingly, we find that in the model-$[1,2,3,4]$, there also exist complex excitations ($\mathsf{E}^c$)  whose geometric structure is quite fruitful and is absent in the X-cube model in 3D.   As mentioned above, the excited state in Fig.~\ref{fig_3b} is obtained by folding the string excitation $(1,2)^{\hat{x}_1,\hat{x}_2}$ at the price of additional energy cost. As a matter of fact, the resulting shape in Fig.~\ref{fig_3b} with both red and yellow lines can be   considered as a complex excitation. It is called ``chairon'' due to its ``chair'' shape. As the chairon example in Fig.~\ref{fig_3b} demonstrates, the most remarkable feature of complex excitations is that the energy is not distributed along manifold-like objects. For instance, when we consider a convergence of yellow and red lines in Fig.~\ref{fig_3b}, it's obvious that the bifurcation of lines can't be homeomorphous to a 1D Euclidean space \cite{EGUCHI1980213}. Therefore, we can't simply label such an excitation by ``string'' or ``loop'', as a result of their non-manifold nature.    Originated from the different flavors of $(1,2)$-type excitations,   chairons can also carry different flavors. By direct calculation, we can find that there are $4 \times \binom{3}{2}=12$ flavors of chairons in total in the model-$[1,2,3,4]$.



Let us focus on the chairon in Fig.~\ref{fig_3b} and concretely carry out the stabilizer operators whose eigenvalues are flipped and then discuss its consequences. For all  $1$-cubes  $\gamma_1$  along the red line within $\hat{x}_1$-$\hat{x}_2$ plane, $B^{\langle\hat{x}_1,\hat{x}_2,\hat{x}_3\rangle}_{\gamma_1}=-1$\,,$B^{\langle\hat{x}_1,\hat{x}_2,\hat{x}_4\rangle}_{\gamma_1}=-1$. For all $1$-cubes $\gamma_1$    along the red line within $\hat{x}_1$-$\hat{x}_3$ plane, $B^{\langle\hat{x}_1,\hat{x}_2,\hat{x}_3\rangle}_{\gamma_1}=-1$\,,$B^{\langle\hat{x}_1,\hat{x}_3,\hat{x}_4\rangle}_{\gamma_1}=-1$. For all $1$-cubes  $\gamma_1$    along the yellow crease line $C^1$, $B^{\langle\hat{x}_1,\hat{x}_2,\hat{x}_4\rangle}_{\gamma_1}=-1$\,,$B^{\langle\hat{x}_1,\hat{x}_3,\hat{x}_4\rangle}_{\gamma_1}=-1$.  these operators form the set $\{\hat{O}\}_{\text{complex}}$.  Since only two operators per $\gamma_1$ along the yellow line are excited, one may conclude that    energy density along the yellow line is still $2K$.  {As a result,   energy is uniformly distributed along both yellow and red lines.}

The mobility and deformability of chairons  are relatively difficult to be described, in contrast to simple excitations where the integer $m$ is good enough. According to our discussion on the mobility and deformability of $(1,2)$-type excitations, the two U-shaped segments (i.e., red lines in Fig.~\ref{fig_3b}) of a chairon are freely deformable within the 2D planes where they are located at without additional energy cost, as long as their lengths stay  the same. But the deformability of the crease line is kind of unspeakable, as it's length and shape are both related to the deformation of the U-shaped segments. As a whole, the chairon can move along the direction of the crease line. However, regarding a chairon as an one-dimensional excitation would be an oversimplification of it's mobility and deformability for sure.

%

\subsubsection{Yuons of 4 flavors}
 
\begin{figure*}[htb]
	\centering  
	\subfigure[A $(1,2)$-type excitation]{
		\label{fig_pad}
		\includegraphics[width=0.24\textwidth]{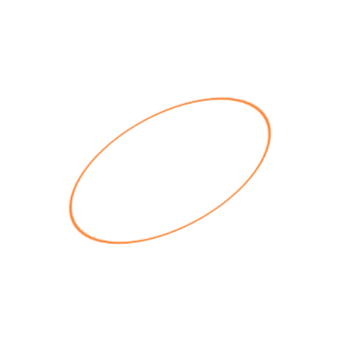}}
	\subfigure[A chairon]{
		\label{fig_chair}
		\includegraphics[width=0.24\textwidth]{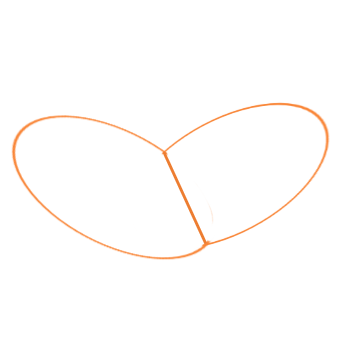}}
	\subfigure[A yuon]{
		\label{fig_yu}
		\includegraphics[width=0.24\textwidth]{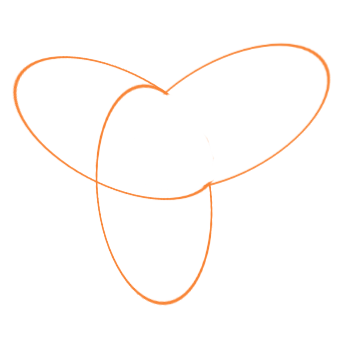}}
	\caption{  Pictorial comparison among a $(1,2)$-type excitation, a chairon and a yuon in the model-$[1,2,3,4]$. }
	\label{fig_com_yuon}
\end{figure*}

Except for chairons, we   also find another kind of complex excitations in the model-$[1,2,3,4]$, which can be dubbed as ``yuon'', since a yuon is a Y-shaped object composed of three U-shaped strings, as shown in Fig.~\ref{fig_com_yuon}. A yuon can be excited by further acting $W(S^2_{III})$ after applying $W(S^2_I)$ and $W(S^2_{II})$ in Eq.~(\ref{eq_S1}) and Eq.~(\ref{eq_S2}), where $S^2_{III}=\{(n+\frac{1}{2}, 0, 0, m+\frac{1}{2})|m,n=0,1,2,\dots, L-1\}$. Although no $B$ term along the convergence line would be excited now, as we've already applied 3 operators on the ground state, three connected U-shaped excitations will still remain, which forms a yuon.  Therefore, the space that can embed a yuon must be at least four dimensional. A schematic comparison among an $(1,2)$-type excitation, a chairon and a yuon in the model-$[1,2,3,4]$ is given in Fig.~\ref{fig_com_yuon}. Similar to chairons, we   find that there are $4$ flavors of yuons in the model-$[1,2,3,4]$.
   
Since chairons and yuons have already covered all kinds of intersection of $(1,2)$-type excitations, we expect these two kinds of excitations can be regarded as the most elementary building blocks for all kinds of complex excitations in the model-$[1,2,3,4]$. In Sec.~\ref{subsec:fractonic}, we will discuss about the model-$[1,2,3,5]$, which has different kinds of ``building blocks''.

\subsection{Excited states with multiple spatially separate loops in the model-$[1,2,3,4]$}\label{sec_multiple_loop}

In the model-$[1,2,3,4]$ discussed here, we may discuss  an excited state  with multiple spatially separate loops each of which has $6$ flavor options. Some typical excited states with two or three loops are listed in Table~\ref{table:comparison_of_fusion_results}. There are several remarks on this table:
\begin{itemize}
\item In pure Abelian topological order, the fusion of two loops doesn't depend on where the two loops are initially located. Case 0 demonstrates the fusion in a pure $\Z_2$ topological order. But in fracton topological order, Table~\ref{table:comparison_of_fusion_results}    shows three different cases for the two-loop fusion due to the (partial) restriction of mobility and deformability in the model-$[1,2,3,4]$. 
\item  In Sec.~\ref{subsec:rev_of_xcube}, we have studied an excited state with  lineons $\ell_x$ and $\ell_y$ in the 3D X-cube model (i.e., the model-$[0,1,2,3]$).  If the two lineons are able to meet at some point, the fusion output is another lineon labeled by $\ell_z$ that can move along the straight line that is along $z$-direction and passes the intersection point. Nevertheless, in the Case 3, two loops that are able to meet can only fuse to a chairon that has a connected non-manifold shape, rather than another new loop.  
\item We will see shortly in Sec.~\ref{subsec:fractonic}, in the model-[$0,1,2,4$], two lineons, if located properly, will fuse into a fracton rather than another lineon. 
\end{itemize}
\begin{table*}[t]
	\centering
	\caption{ Typical examples of excited states with spatially separate loops in the model-$[1,2,3,4]$.   The two loops labeled by (*) are in the same $\hat{x}_1$-$\hat{x}_2$ plane; The two loops labeled by (**) are in different $\hat{x}_1$-$\hat{x}_2$ planes. Case 1 to Case 4 are two-loop states in the model-$[1,2,3,4]$, while Case 5 is a three-loop state. ``Case 0'' is added for comparison, which  happens in pure $\Z_2$ topological order.}
	\label{table:comparison_of_fusion_results} 
	\begin{tabular}{cccccc}
		\hline
		
		\hline
		&1st loop & 2nd loop & 3rd loop & Sectors & Fusion output \\		\hline

 	\textit{Case 0}&	any position & \begin{minipage}{1.2in}any position\end{minipage}&  $\times$ & $\mathbb{I}$  & vacuum   \\			 
 	\textit{Case 1}&	$(1,2)^{\hat{x}_1,\hat{x}_2}$ & \begin{minipage}{1.2in}$(1,2)^{\hat{x}_1,\hat{x}_2}$ (*) \end{minipage}&  $\times$ & $\mathbb{I}$  & vacuum   \\			 

	\textit{Case 2}&			$(1,2)^{\hat{x}_1,\hat{x}_2}$ & \begin{minipage}{1.2in}$(1,2)^{\hat{x}_1,\hat{x}_2}$(**)    \end{minipage} &  $\times$ & $\mathsf{E}^d$  &  intrinsically disconnected\\

	\textit{Case 3}&	\begin{minipage}{0.7in}	~\\$(1,2)^{\hat{x}_1,\hat{x}_2}$ \end{minipage}& \begin{minipage}{1in}	~\\ $(1,2)^{\hat{x}_1,\hat{x}_3}$ \end{minipage}& $\times$ &  \begin{minipage}{0.5in}$\mathsf{E}^c$\end{minipage} &$\vtop{\vskip-13pt\hbox{\includegraphics[width=0.25\textwidth]{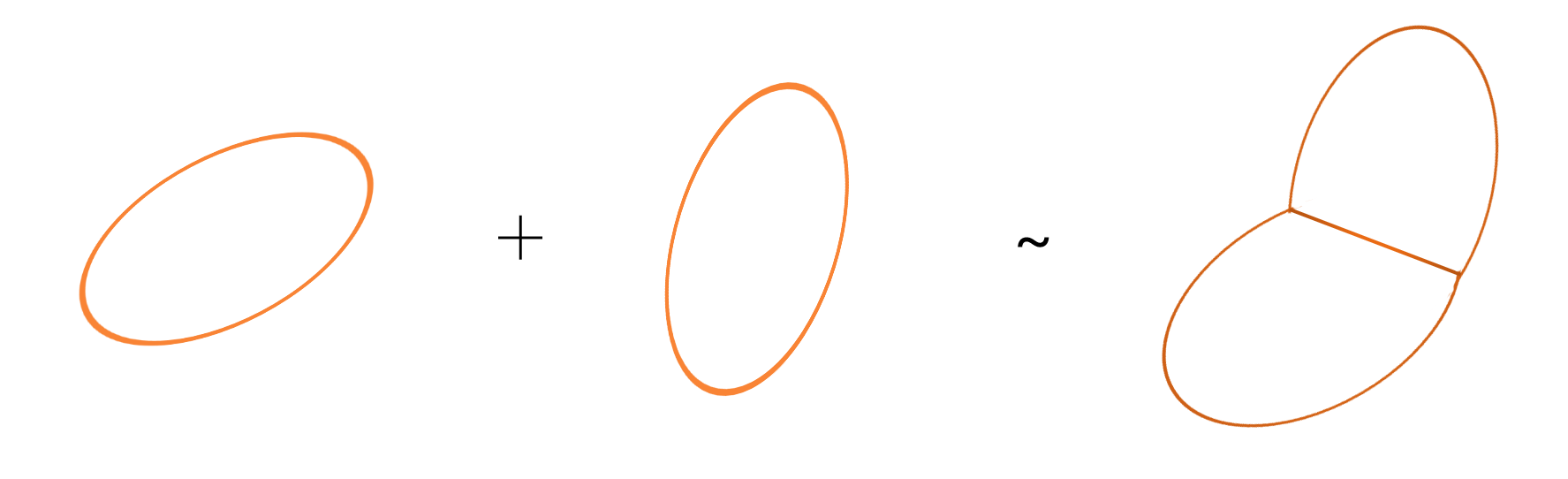}}}$  : chairon \\			 

 			\textit{Case 4}&$(1,2)^{\hat{x}_1,\hat{x}_2}$ & $(1,2)^{\hat{x}_3,\hat{x}_4}$ & $\times$ &  $\mathsf{E}^d$  & intrinsically disconnected \\

	 \textit{Case 5}&		\begin{minipage}{0.7in}$(1,2)^{\hat{x}_1,\hat{x}_2}$ \end{minipage}& \begin{minipage}{1in}$(1,2)^{\hat{x}_1,\hat{x}_3}$ \end{minipage}& \begin{minipage}{1in}~$(1,2)^{\hat{x}_1,\hat{x}_4}$ \end{minipage}&\begin{minipage}{0.5in}$\mathsf{E}^c$ \end{minipage}& $\vtop{\vskip-10pt\hbox{\includegraphics[width=0.25\textwidth]{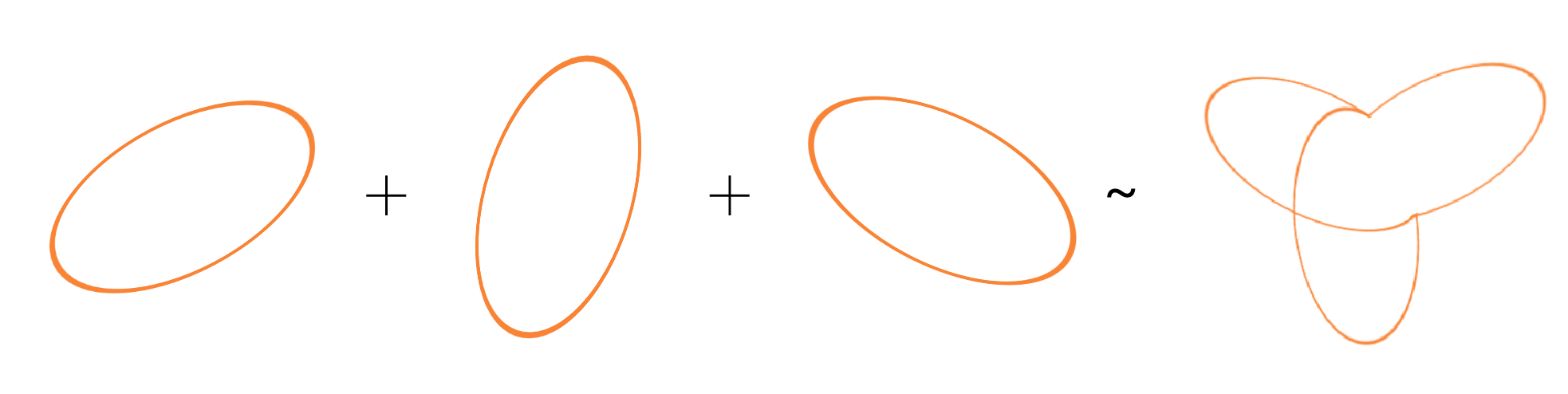}}}$ : yuon \\		\hline

		\hline          
	\end{tabular}
\end{table*}

\subsection{Gravity analog: energy density and spacetime curvature}
\label{subsec:ex_1234series}

In lattice of arbitrary dimension higher than $3$, as our argument doesn't rely on specific dimensions, we would expect our results still persist. That is to say, a $[D-3,D-2,D-1,D]$ model would contain $(0,0)$-type, $(0,D-1)$-type and $(D-3,D-2)$-type excitations. More excitations in $[D-3,D-2,D-1,D]$ models for $D=3,4,5$ are listed in Table.~\ref{table:excitations}. Moreover, in higher dimensional cases, there are  similar situations that when we act $F^{D-2}(C^{D-3})=\prod\limits_{{\gamma_{D-2}}\in S^{D-2}_{I}} \sigma^x_{\gamma_{D-2}} \prod\limits_{{\gamma_{D-2}}\in S^{D-2}_{II}} \sigma^x_{\gamma_{D-2}}$, where $S^{D-2}_{I} \cap S^{D-2}_{II} = \partial S^{D-2}_{I} \cap \partial S^{D-2}_{II} = C^{D-3}$, $B^l_{\gamma_{D-3}}$ terms along $C^{D-3}$ will be excited for certain $l$'s. Such a phenomenon naturally reminds us of gravity, considering that there are already some works concerning about this problem.  \cite{Pretko2018,Pretko2017}  Especially, in the $6$-dimensional model $[3,4,5,6]$, when the scale considered is much larger than the lattice constant, we would see condensations of flat closed $4$-manifolds in the ground state. When we gradually heat up the system, energy density will rise   where the $4$-manifolds curve. Nevertheless, despite the direct correspondence between curvature and energy density, the curvature which matters here is extrinsic curvature, while in general relativity the correspondence is between intrinsic curvature and stress-energy tensor. As a result, the relation between our lattice models and gravity is still vague.

\section{General construction of lattice models}
\label{sec:family}

\subsection{Lattice Hamiltonians}
\label{subsec:cons_family}

As our previous sections demonstrated, we can promote $(0,1)$-type excitations to $(i,i+1)$-type excitations by lifting the dimensions of all the $n$-cubes where spins and other operators are located at by $i$. Naturally, one may be curious about, if it's possible to define spins and operators on different kinds of $n$-cubes, without any redundant constraints? To deal with this problem, we come up with a further generalization procedure. In this procedure,   the dimension of the objects on which the operators and spins are defined, can be adjusted independently. Since we are focusing on fracton models, other than the dimensions of spins, lower-dimensional cube operators, higher-dimensional cube operators and the total space, we also need the dimension of leaf spaces to specify such a model. For instance, for the X-cube model, we have spin dimension $d_s=1$, lower-dimensional cube operator dimension $d_n=0$, higher-dimensional cube operator dimension $d_c=3$, space dimension $D=3$ and leaf dimension $d_l=2$. Generally, it seems that we need five dimension indexes $[d_n,d_s,d_l,d_c,D]$ to specify a member in the model ``family''. Given such a 5-tuple $[d_n,d_s,d_l,d_c,D]$, the Hamiltonian of the corresponding member model is:
\begin{align}
\label{eq:the_Hamiltonian}
H_{[d_n,d_s,d_l,d_c,D]} = -J\sum_{\{{\gamma}_{d_c}\}} A_{{\gamma}_{d_c}} - K \sum_{\{{\gamma}_{d_n}\}} \sum_{l} B^l_{{\gamma}_{d_n}},
\end{align}
where the definition of the terms is given below:

\begin{itemize}
	\item A $B^l_{{\gamma}_{d_n}}$ term is the product of $z$-components of the $2^{d_s-d_n} \binom{d_l-d_n}{d_s-d_n}$ spins whose coordinates  are obtained by shifting $(d_s-d_n)$ coordinates of $\gamma_{d_n}$ along the directions in $\mathcal{C}^l_{\gamma_{d_n}}$ by $\pm \frac{1}{2}$. Here $\mathcal{C}^l_{\gamma_{d_n}} \equiv \mathcal{L} \cap \mathcal{C}^i_{\gamma_{d_n}}$.
	\item An $A_{{\gamma}_{d_c}}$ term is the product of $x$-components of the $2^{d_c-d_s}\binom{d_c}{d_c-d_s}$ spins whose coordinates  are obtained by shifting $(d_c-d_s)$ coordinates of $\gamma_{d_c}$ along the directions in $\mathcal{C}^h_{{\gamma}_{d_c}}$ by $\pm \frac{1}{2}$.
\end{itemize}

Take the model-$[1,2,3,4]$ as an example. For a given $\gamma_1=(0,0,0,\frac{1}{2})$, we can see that $\mathcal{C}^i_{(0,0,0,\frac{1}{2})}=\{{\hat{x}_1},{\hat{x}_2},{\hat{x}_3}\}$, while $\mathcal{C}^h_{(0,0,0,\frac{1}{2})}=\{{\hat{x}_4}\}$. For the 3 leaf spaces $\langle {\hat{x}_1},{\hat{x}_2},{\hat{x}_4}\rangle$, $\langle {\hat{x}_1},{\hat{x}_3},{\hat{x}_4}\rangle$ and $\langle {\hat{x}_2},{\hat{x}_3},{\hat{x}_4}\rangle$ associated with $(0,0,0,\frac{1}{2})$, we have $\mathcal{C}^{\langle {\hat{x}_1}, {\hat{x}_2}, {\hat{x}_4}\rangle}_{(0,0,0,\frac{1}{2})}=\{{\hat{x}_1}, {\hat{x}_2}\}$, $\mathcal{C}^{\langle {\hat{x}_1}, {\hat{x}_3}, {\hat{x}_4}\rangle}_{(0,0,0,\frac{1}{2})}=\{{\hat{x}_1},{\hat{x}_3}\}$ and $\mathcal{C}^{\langle {\hat{x}_2}, {\hat{x}_3}, {\hat{x}_4}\rangle}_{(0,0,0,\frac{1}{2})}=\{{\hat{x}_2},{\hat{x}_3}\}$ respectively, so we can obtain the $B^l_{(0,0,0,\frac{1}{2})}$ as in Eq.~(\ref{eq:B terms}). Similarly, for the 4-cube $(\frac{1}{2},\frac{1}{2},\frac{1}{2},\frac{1}{2})$, $A_{(\frac{1}{2},\frac{1}{2},\frac{1}{2},\frac{1}{2})}$ can be simply obtained as in Eq.~(\ref{eq:A terms}).

In the following part of this section, we will use $\mathcal{S}_c$ to refer to the set of the nearest spins of the $d_c$-cube $c$, $\mathcal{S}_n$ to refer to the set of the nearest spins of the $d_n$-cube $n$, and $\mathcal{S}_n^l$ to refer to the set of the nearest spins of the $d_n$-cube $n$ inside the $d_l$-dimensional leaf space $l$ (here $l$ is associated with $n$). Apparently, then we have $\mathop{\cup}\limits_{l} \mathcal{S}^l_n = \mathcal{S}_n $.

Though we are trying to make the choice of different dimension indexes independent to each other, we still need to respect some orders of the dimensions. Firstly, we find that $d_n$ and $d_c$ can't be equal to $d_s$, otherwise the cube operators would be trivialized. Besides, according to the dimension order of X-cube model, we expect $d_n$ to be smaller than $d_s$ while $d_c$ should be larger than $d_s$. Furthermore, $d_l>d_s$, $d_l<D$ and $d_c \leq D$ are obviously required. However, it should be noted that the condition $d_s$  is between $d_c$ and $d_n$ is not really necessary in defining an exactly solvable fracton order model. For simplicity, we will focus on cases where the condition is satisfied in this article.

Since we expect our models to be exactly solvable, we require every higher dimensional cube operator shares even or zero number of nearest spins with any lower dimensional cube operator. Since lower dimensional cube operators are embedded in different leaf spaces, this condition means that  
\begin{align}
\left| \mathcal{S}^l_n \cap \mathcal{S}_c \right| \;\text{mod}\; 2=0 \; \forall l,n,c.
\end{align}

Acccording to the symmetry of the cubic lattice, we can calculate $\left| \mathcal{S}^l_n \cap \mathcal{S}_c \right|$ for any pair of nearest $\gamma_{d_n}$ and $\gamma_{d_c}$ in the lattice. Therefore, we only need to consider the number of spins shared by $\gamma_{d_c}$ $c_1=(\underbrace{\frac{1}{2}, \frac{1}{2},\dots, \frac{1}{2}}_{d_c},\underbrace{0,0,\dots,0}_{D-d_c})$ and $\gamma_{d_n}$ $n_1=(\underbrace{\frac{1}{2}, \frac{1}{2},\dots, \frac{1}{2}}_{d_n},\underbrace{0,0,\dots,0}_{D-d_n})$. Apparently, for a spin $s_1$ that is nearest to both $c_1$ and $n_1$, the first $d_n$ coordinates of $s_1$ must be $\frac{1}{2}$ and the last $(D-d_c)$ coordinates must be $0$, only the values of the $(d_c-d_n)$ coordinates in the middle are variable. 

To calculate $\left| \mathcal{S}^l_n \cap \mathcal{S}_c \right|$, we only need to care about the uncertain middle part of the coordinates of $s_1$, which is composed of $(d_c-d_n)$ numbers. Each subsequence consists of these $(d_c-d_n)$ numbers with $(d_s-d_n)$ digits being $\frac{1}{2}$ and the others being $0$ corresponds to a spin which is simultaneously nearest to $c_1$ and $n_1$. As a result, a shared spin will take the form $s_1=(\underbrace{\frac{1}{2}, \frac{1}{2},\dots, \frac{1}{2}}_{d_n},\underbrace{\dots}_{d_c-d_n},\underbrace{0,0,\dots,0}_{D-d_c})$. For leaf $l$ associated with $n_1$, which contains $\alpha$ directions with uncertain coordinates (i.e. in the middle part of $s_1$), we have $\left| \mathcal{S}^l_{n_1} \cap \mathcal{S}_{c_1} \right|=\tbinom{\alpha}{d_s-d_n}$. However, as we expect the parity of $\left| \mathcal{S}^l_{n_1} \cap \mathcal{S}_{c_1} \right|$ to be independent of the choice of leaf, $\alpha$ should be insensitive to the choice of leaf space, which means all leaves must have the same number of uncertain digits (here we ignore the case where the change of $\alpha$ doesn't influence the parity of $\left| \mathcal{S}^l_{n_1} \cap \mathcal{S}_{c_1} \right|$ for simplicity). Therefore, the last part of sequence $s_1=(\underbrace{\frac{1}{2}, \frac{1}{2},\dots, \frac{1}{2}}_{d_n},\underbrace{\dots}_{d_c-d_n},\underbrace{0,0,\dots,0}_{D-d_c})$ must vanish, i.e. $D-d_c$ must be 0. Then we have $\alpha =d_l-d_n \; \forall \ l$. And the exactly solvable condition is just
\begin{align}
\left| \mathcal{S}^l_n \cap \mathcal{S}_c \right| \; \text{mod} \; 2=\tbinom{d_l-d_n}{d_s-d_n} \; \text{mod} \; 2=0
\end{align}
together with
\begin{align}
d_n<d_s<d_l<d_c=D.
\end{align}

Since then, we only need a $4$-tuple $[d_n,d_s,d_l,d_c]$ (or $[d_n,d_s,d_l,D]$) to specify an exactly solvable model.

\subsection{Family tree}
\label{subsec:tree}

Based on our $4$-tuple notation of models, we can understand the actual meaning of the label ``$[0,1,2,3]$'' of   X-cube model. With such a notation manner, we can easily obtain that X-cube is the simplest model in this series. As a result, we can use it as the starting point of a ``family tree'' of the generalized models, which is depicted in Fig.~\ref{fig_tree}. As an example of the novel properties of the models on the tree, we will demonstrate that there are new kinds of complex excitations in $[D-4,D-3,D-2,D]$-type of models in the next subsection. Here we would like to give a preliminary description of the ground states and energy spectrum of the models on the tree.

Because the Hamiltonian of a general model is similar to $[D-3,D-2,D-1,D]$ model, which is given in Eq.~(\ref{eq:1234_branch_ham}), the ground states of a general model will obey a set of conditions of the following form:

\begin{align}
\label{eq:super ground condition}
A_{{\gamma}_{d_c}} \ket{\phi} = \ket{\phi},\ B^l_{{\gamma}_{d_n}} \ket{\phi} = \ket{\phi}\ \forall\ {\gamma}_{d_c}, \ {\gamma}_{d-n},\ l.
\end{align}

As always, with the $\sigma_z$-basis, we can see that every configuration is an eigenvector of an arbitrary $B$ operator, and the total Hilbert space can be spanned by all the configurations. Furthermore, the $B$ conditions in Eq.~(\ref{eq:super ground condition}) require the eigenvalue of any $B$ operator for all configurations in a ground state to be $1$. That is to say, for any $\gamma_{d_n}$ in a ground state configuration, either of the following conditions must be satisfied:

\begin{itemize}
	\item No nearest spin is altered;
	\item For each pair of nearest spins linked by the $\gamma_{d_n}$, there is exactly one spin of the pair being altered (for models where $d_s-d_n \geq 2$ a ``pair'' should be promoted to a set of $2^{d_s-d_n}$ spins).
\end{itemize}

Then, the $A$ condition in Eq.~(\ref{eq:super ground condition}) can be seen as requiring all the ground state configurations which can be transformed to each other by acting $A$ operators share the same weight. Therefore, we can find that the unique ground state of a general model with open boundary conditions is $\ket{\Phi}=\prod\limits_{\gamma_{d_c}} \frac{1+A_{\gamma_{d_c}}}{\sqrt{2}} \ket{\uparrow \uparrow \uparrow\dots\uparrow}$, where $\ket{\uparrow \uparrow \uparrow\dots\uparrow}$ refers to the reference state. Besides, please note that here the form of $A_{\gamma_{d_c}}$ also implicitly depends on $d_s$. With periodic boundary condition, the ground states of the tree models are expected to be degenerate. We've found signs that suggest the ground state degeneracy of these models may be more complicated that the known subextensive growth. Relevant results will be involved in our future work Ref.~\cite{next_episode}.

For models on the family tree (see Fig.~\ref{fig_tree}), all simple excitations can be classified into two classes: $(d_n,d_s)$-type excitations and $(0,0)$-type excitations. Moreover, since segments of complex excitations can be regarded as the convergence of several $(d_n,d_s)$-type excitations, and the energy density along the segments can be determined by the number of converged $(d_n,d_s)$-type excitations, we only need to consider the energy cost of such convergences to determine the energy cost of a complex excitation. As in Sec.~\ref{subsec:ex_1234}, here we can conclude the data of the most important simple excitations in a general model as below:

\begin{itemize}
	\item $A_{\gamma_{d_c}}=-1$ excitations, $(0,0)$-type, generated by $\prod\limits_{{\gamma_{d_s}}\in D^{d_c-d_s}} \sigma^z_{\gamma_{d_s}}$. The excitations sit on the vertices of the $D^{d_c-d_s}$. 
	\item $B^l_{\gamma_{d_n}}=-1$ excitations, $(d_n,d_s)$-type, generated by $\prod\limits_{{\gamma_{d_s}}\in S^{d_s}} \sigma^x_{\gamma_{d_s}}$. The excitations sit on the boundary of $S^{d_s}$. 
\end{itemize}

As for the energy cost, simply we can find that the energy cost of such a $(0,0)$-type excitation is always $J$, so the energy cost of different  {groups of fractons} are respectively $2J$, $4J$, \dots, $2^{d_c-d_s-1}J$. Most of the  {groups} are fractons, except for the last one which are $(0,d_s+1)$-type excitations (or disconnected excitations). 

The spectrum of convergences of $(d_n,d_s)$-type excitations (i.e. segments of complex excitations) is more difficult to calculate. Generally, for a specific $\gamma_{d_n}$ $n_1$, we can find that the number of excited $B_{n_1}^l$ operators is determined by the number of pairs of spins around the $n_1$ which contain exactly one altered spin. For simplicity, such a pair will be regarded as ``excited''. Therefore, we can label a convergence of $(d_n,d_s)$-type excitations at $n_1$ as an $i$-convergence, where $i$ is the number of excited pairs linked by $n_1$. As an $i$-convergence always has the same energy as a $(\binom{d_l-d_n}{d_s-d_n}-i)$-convergence, we only need to consider $i \leq \lfloor \frac{\binom{d_l-d_n}{d_s-d_n}}{2} \rfloor$ in this article. Since there are $\binom{d_c-d_n}{d_s-d_n}$ pairs of spins linked by a given $\gamma_{d_n}$, and a leaf always contains $\binom{d_l-d_n}{d_s-d_n}$ such pairs, we can see that the energy cost of an $i$-convergence is just the number of different combinations of $\binom{d_l-d_n}{d_s-d_n}$ pairs (i.e. leaves) which contain odd number of excited pairs. For a given $i$ we can find that there are $\sum^{\lfloor \frac{i-1}{2} \rfloor}_{x=0}\binom{i}{2x+1}\binom{\binom{d_c-d_n}{d_s-d_n}-i}{\binom{d_l-d_n}{d_s-d_n}-2x-1}$ such combinations, so the energy cost of an $i$-convergence on a $\gamma_{d_n}$ is:

\begin{align}
E_i = \sum^{\lfloor \frac{i-1}{2} \rfloor}_{x=0}\binom{i}{2x+1}\binom{\binom{d_c-d_n}{d_s-d_n}-i}{\binom{d_l-d_n}{d_s-d_n}-2x-1} K.
\end{align}

For instance, we can consider the the model-$[1,2,3,5]$. Since $\frac{\binom{d_l-d_n}{d_s-d_n}}{2}=\frac{5}{2} \geq 2$, there is only one kind of convergences of $(1,2)$-type excitations need to be discussed, that is the $2$-convergences. For instance, $(\frac{1}{2},0,0,0,0)$ links 4 pairs of spins, $(\frac{1}{2},\pm \frac{1}{2},0,0,0)$, $(\frac{1}{2},0,\pm \frac{1}{2},0,0)$, $(\frac{1}{2},0,0,\pm \frac{1}{2},0)$ and $(\frac{1}{2},0,0,0,\pm \frac{1}{2})$, while a leaf like $\langle x_1, x_2, x_3\rangle$ contains two of such pairs. So we can find that of the $\binom{5-1}{3-1}=6$ kinds of possible combinations of pairs, there are $\sum^{\lfloor \frac{2-1}{2} \rfloor}_{x=0}\binom{2}{2x+1}\binom{\binom{5-1}{2-1}-2}{\binom{3-1}{2-1}-2x-1}=4$ combinations that contain odd number of excited pairs, so there are 4 $B^l_{(\frac{1}{2},0,0,0,0)}$ terms being excited. Such a 2-convergence can exist as a segment of a complex excitation ($\beta$-chairon, see Sec.~\ref{subsec:fractonic}) in the model-$[1,2,3,5]$, and its energy cost is proportional to its length. More excitations in the model-$[1,2,3,5]$ is given in Table.~\ref{table:1235_excitations}.

\subsection{Simple excitations in the model-$[0,1,2,4]$ and $[1,2,3,5]$}
\label{subsec:fractonic}

In this subsection, we will take the two models ``$[0,1,2,4]$'' and ``$[1,2,3,5]$'' on the family tree to exemplify the novel properties of models outside the $[D-3,D-2,D-1,D]$ series. 

Let's start with  the model-$[0,1,2,4]$. It's easy to check that, unlike the model-$[1,2,3,4]$, the model-$[0,1,2,4]$ doesn't contain any spatially extended excitations, which makes its spectrum much simpler. As in the X-cube model, $(0,1)$-type excitations, i.e., lineons, are generated at the ends of straight string operators consisted of $\sigma^x$'s, and fractons are generated at the corners of cube (i.e. $D^3$) operators consisted of $\sigma^z$'s. However, there is indeed something exotic in the model-$[0,1,2,4]$: for example, since the model is defined on a 4-dimensional lattice, the convergence of two straight strings can only dual to another convergence, so the 2-convergence becomes a fracton. While in X-cube model, since there is a duality between plaquettes and links, the point-like excitation at such a convergence can be moved along the line perpendicular to the convergence. In some sense, due to the higher space dimension, a kind of lineons in $[0,1,2,3]$ model are frozen in the model-$[0,1,2,4]$.  Let us  point out some key properties of lineons and fractons in the model-$[0,1,2,4]$ summarized in Table~\ref{table:0124_excitations}:
\begin{itemize}
\item All topological non-trivial excitations in $[0,1,2,4]$ are point-like (here excited states composed of discrete points are also recognized ``point-like''), belonging to $\mathsf{E}^s$ or $\mathsf{E}^d$ sectors.
\item In contrast to $[0,1,2,3]$, fractons can be formed by either flipping $A$ or $B$ stabilizers. There are one kind of fractons labeled by $A$, but there are six kinds of fractons labeled by $B$ stabilizers due to the six different leaf space indices $\ell$\footnote{See the ``Example 2'' in Sec.~\ref{sec_pre_leaf}.}.
\item \emph{Two lineons can fuse into a fracton.} There are four types of lineons that can move along parallel straight lines of $\hat{x}_1,\hat{x}_2,\hat{x}_3,\hat{x}_4$ orthogonal directions. Picking two different types of lineons from four (totally 6 choices), if  the two straight lines where the two lineons can move intersect at some point, the two lineons fuse into a fracton with flipped stabilizers $B^l_{\gamma_0}$ where leaf space indices $\ell$ exactly have $6$ corresponding choices. In Sec.~\ref{subsec:rev_of_xcube}, we have reviewed that, in the model-$[0,1,2,3]$, the fusion output of two lineons if they can meet from orthogonal directions is, however, another lineon.
\end{itemize}
\begin{table*}[htp] 	
	\caption{Typical examples of  excitations in the model-$[0,1,2,4]$.   }
	
	\begin{tabular}{ccccc}
		\hline
		
		\hline
		Excitations &~~~Sectors~~~& Flipped stabilizers & Creation operators \\
		\hline
		fracton: {$(0,0)$} & $\mathsf{E}^s$ & $A_{\gamma_4}$ & $\prod\limits_{{\gamma}_1 \in D^3} \sigma^z_{{\gamma}_1}$ 
		\\
		lineon: $(0,1)$ & $\mathsf{E}^s$ & $B^l_{\gamma_0}$ & $\prod\limits_{{\gamma}_1 \in S^1} \sigma^x_{{\gamma}_1}$
		\\
		connected planeon: $(0,2)$ & $\mathsf{E}^s$ & $A_{\gamma_4}$ & $\prod\limits_{{\gamma}_1 \in D^1_1}\sigma^z_{{\gamma}_1}$
		\\
	fracton:	 {$(0,0)$} & $\mathsf{E}^s$ & $B^l_{\gamma_0}$ & $\prod\limits_{{\gamma}_1 \in S^1_{I}} \sigma^x_{{\gamma}_1} \prod\limits_{{\gamma}_1 \in S^1_{II}} \sigma^x_{{\gamma}_1}$, where $\partial S^1_{I} \cap \partial S^1_{II} = C^0 \neq \emptyset$
		\\
		 
disconnected planeon  & $\mathsf{E}^d$ & $A_{\gamma_4}$ & $\prod\limits_{{\gamma}_1 \in D^3} \sigma^z_{{\gamma}_1}$  
		\\
		
		\hline
		
		\hline	
	\end{tabular}
	\label{table:0124_excitations}
\end{table*}

Then we consider the the model-$[1,2,3,5]$. The Hamiltonian of $[1,2,3,5]$ is:
\begin{align}
H_{[1,2,3,5]} = -J\sum_{\{{\gamma}_{5}\}} A_{{\gamma}_{5}} - K \sum_{\{{\gamma}_{1}\}} \sum_{l} B^l_{{\gamma}_{1}}.
\end{align}

The simple excitations in the model-$[1,2,3,5]$ are mostly the same as in $[1,2,3,4]$, except the mobility of  {groups} of fractons. As basic fractons are located at the vertices of $D^3$'s now, pairs of fractons at the vertices of $D^2_2$'s become $(0,0)$-type excitations, while the tetrads of fractons at the vertices of $D^1_2$'s belong to $(0,3)$-type.\footnote{See the ``Example 3'' in Sec.~\ref{sec_pre_leaf} for an introduction to the leaves in model-$[1,2,3,5]$.} 

Similar to Sec.~\ref{subsec:ex_1234}, here we can classify the $(1,2)$-type excitations $W(S^2)$ in the model-$[1,2,3,5]$ into $\binom{5}{2}=10$ flavors, according to the plane (i.e. $\hat{x}_1$-$\hat{x}_2$, $\hat{x}_1$-$\hat{x}_3$, $\hat{x}_1$-$\hat{x}_4$, $\hat{x}_1$-$\hat{x}_5$, $\hat{x}_2$-$\hat{x}_3$, $\hat{x}_2$-$\hat{x}_4$, $\hat{x}_2$-$\hat{x}_5$, $\hat{x}_3$-$\hat{x}_4$, $\hat{x}_3$-$\hat{x}_5$, $\hat{x}_4$-$\hat{x}_5$ ) where $S^2$ is located at. Generally, when we act a $W(S^2)$ on the ground state, where $S^2$ is located at a $\hat{x}_i$-$\hat{x}_j$ plane, then at a $\gamma_1$ along the boundary of $S^2$, eigenvalues of $B^{\langle \hat{x}_i, \hat{x}_j,\hat{x}_k \rangle}_{\gamma_1}$, $B^{\langle \hat{x}_i, \hat{x}_j,\hat{x}_h \rangle}_{\gamma_1}$ and $B^{\langle \hat{x}_i, \hat{x}_j,\hat{x}_p \rangle}_{\gamma_1}$ will be flipped. Here $i,j,h,k,p \in \{1,2,3,4,5\}$, and $i,j,h,k,p$ are all different from each other. For instance, by acting $W(S^2_I)$ on the ground state, where $S^2_I=\{(n+\frac{1}{2}, m+\frac{1}{2}, 0, 0, 0)|m,n=0,1,2,\dots, L-1\}$, for an arbitrary $\gamma_1$ along the boundary of $S^2_I$, the eigenvalue of $B^{\langle \hat{x}_1, \hat{x}_2,\hat{x}_3 \rangle}_{\gamma_1}$, $B^{\langle \hat{x}_1, \hat{x}_2,\hat{x}_4 \rangle}_{\gamma_1}$ and $B^{\langle \hat{x}_1, \hat{x}_2,\hat{x}_5 \rangle}_{\gamma_1}$ will be flipped. As a result, we obtain that the energy cost of a string excitation in the model-$[1,2,3,5]$ is $3KL$, where $L$ is the length of the string. More information of the excitations in the family tree models are summarized in Table.~\ref{table:0124_excitations} and Table.~\ref{table:1235_excitations}.

\begin{figure}[t]
	\includegraphics[width=0.3\textwidth]{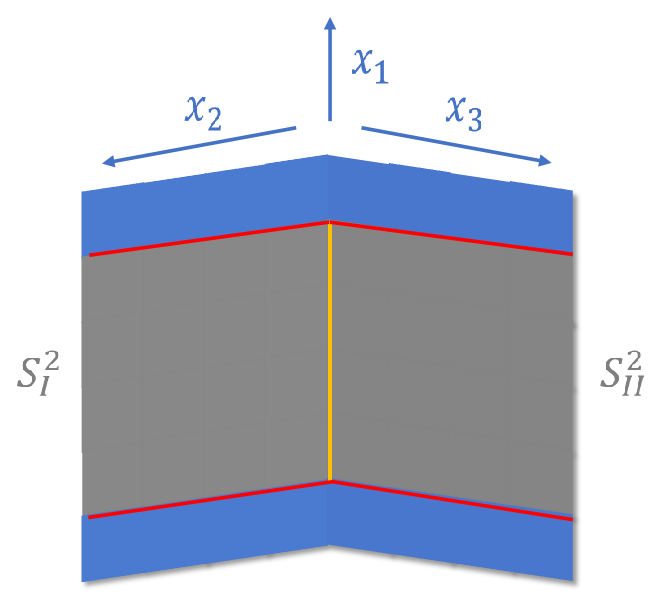}
	\caption{   A $\beta$-chairon in the model-${[1,2,3,5]}$. As in Fig.~\ref{fig_2}, here the grey plaquettes refer to spins on which a $W(S^2)$ is applied. The energy density along red lines is lower than the yellow line, according to our analysis in Sec.~\ref{subsec:com_1235}. That is to say, energy is ununiformly distributed for a $\beta$-chairon.}
	\label{fig_(1,0)}
\end{figure}

\subsection{Complex excitations in the model-$[1,2,3,5]$ (``chairon'', ``cloverion'' and ``xuon'')}

\label{subsec:com_1235}

\begin{table*}[htp] 	
	\caption{Typical examples of excitations in the model-$[1,2,3,5]$. }
	
	\begin{tabular}{ccccc}
		\hline
		
		\hline
		Excitations &~~Sectors~~& Flipped stabilizers & Creation operators \\
		\hline
		 {$(0,0)$} & $\mathsf{E}^s$ & $A_{\gamma_5}$ & $\prod\limits_{{\gamma}_2 \in D^3} \sigma^z_{{\gamma}_2}$ 
		\\
		$(1,2)$ & $\mathsf{E}^s$ & $B^l_{\gamma_1}$ & $\prod\limits_{{\gamma}_2 \in S^2} \sigma^x_{{\gamma}_2}$
		\\
		\begin{minipage}{1.5in} connected  volumeon $(0,3)$ \end{minipage}&$\mathsf{E}^s$ & $A_{\gamma_5}$ & $\prod\limits_{{\gamma}_2 \in D^1_2}\sigma^z_{{\gamma}_2}$
		\\
		 
		$\beta$-Chairon & $\mathsf{E}^c$ & $B^l_{\gamma_1}$ & $\prod_{{\gamma_2}\in S^2_I} \sigma^x_{\gamma_2} \prod_{{\gamma_2}\in S^2_{II}} \sigma^x_{\gamma_2}$, where $\partial S^2_{I} \cap \partial S^2_{II} = C^1 \neq \emptyset$
		\\\\
		Cloverion & $\mathsf{E}^c$ & $B^l_{\gamma_1}$ & \begin{minipage}{3in} $\prod_{{\gamma_2}\in S^2_I} \sigma^x_{\gamma_2} \prod_{{\gamma_2}\in S^2_{II}} \sigma^x_{\gamma_2} \prod_{{\gamma_2}\in S^2_{III}} \sigma^x_{\gamma_2}$, where $\partial S^2_{I} \cap \partial S^2_{II} \cap \partial S^2_{III} = C^1 \neq \emptyset$\end{minipage}
		\\\\
		Xuon & $\mathsf{E}^c$ & $B^l_{\gamma_1}$ & \begin{minipage}{3in}  $\prod_{{\gamma_2}\in S^2_I} \sigma^x_{\gamma_2} \prod_{{\gamma_2}\in S^2_{II}} \sigma^x_{\gamma_2} \prod_{{\gamma_2}\in S^2_{III}} \sigma^x_{\gamma_2} \prod_{{\gamma_2}\in S^2_{IV}} \sigma^x_{\gamma_2}$, where $\partial S^2_{I} \cap \partial S^2_{II} \cap \partial S^2_{III} \cap \partial S^2_{IV}= C^1 \neq \emptyset$\end{minipage}
		\\
		 
	\begin{minipage}{1.3in}	disconnected   volumeon \end{minipage}& $\mathsf{E}^d$ & $A_{\gamma_5}$ & $\prod\limits_{{\gamma}_2 \in D^3} \sigma^z_{{\gamma}_2}$ 
		\\
		
		\hline
		
		\hline	
	\end{tabular}
	\label{table:1235_excitations}
\end{table*}

As in the model-$[1,2,3,4]$, we can find a series of complex excitations as building blocks of general complex excitations in the model-$[1,2,3,5]$. Some typical examples are collected in Table~\ref{table:1235_excitations}. At first, if we further apply $W(S^2_{II})$ after $W(S^2_I)$ ($W(S^2_I)$ is given in the previous subsection), where $S^2_{II}=\{(n+\frac{1}{2}, 0,  m+\frac{1}{2}, 0, 0)|m,n=0,1,2,\dots, L-1\}$, we will have a chairon excitation, which is schematicly presented in Fig.~\ref{fig_(1,0)}. Though the shape of the chairon is the same as in the model-$[1,2,3,4]$, at $(\frac{1}{2},0,0,0,0) \in \partial S^2 \cap \partial S^2_{II}$, now we have $B^{\langle \hat{x}_1, \hat{x}_2,\hat{x}_4 \rangle}_{(\frac{1}{2},0,0,0,0)}$, $B^{\langle \hat{x}_1, \hat{x}_2,\hat{x}_5 \rangle}_{(\frac{1}{2},0,0,0,0)}$, $B^{\langle \hat{x}_1, \hat{x}_3,\hat{x}_4 \rangle}_{(\frac{1}{2},0,0,0,0)}$ and $B^{\langle \hat{x}_1, \hat{x}_3,\hat{x}_5 \rangle}_{(\frac{1}{2},0,0,0,0)}$ being flipped. As now there are $4$ flipped $B$ terms at $(\frac{1}{2},0,0,0,0)$, we can find that unlike in the model-$[1,2,3,4]$ or pure topological order, here the energy is distributed along the excitation unevenly. As a result, we can name the chairon in the model-$[1,2,3,4]$ as $\alpha$-chairon, and the chairon in the model-$[1,2,3,5]$ as $\beta$-chairon, to stress their different distributions of energy. More generally, different types of chairons can be distinguished by the number of flipped stabilizers along the excitation: if the number is a constant, then we call it an $\alpha$-chairon. Otherwise, it's a $\beta$-chairon. Again, by counting the possible combinations of different dimensions, we find that there are $\binom{5}{1} \times \binom{4}{2}=30$ flavors of $\beta$-chairons in the model-$[1,2,3,5]$.

Furthermore, by acting $W(S^2_{III})$ after $W(S^2_{I})$ and $W(S^2_{II})$, where $S^2_{III}=\{(n+\frac{1}{2}, 0, 0, m+\frac{1}{2}, 0)|m,n=0,1,2,\dots, L-1\}$. For $\gamma_1=(\frac{1}{2},0,0,0,0) \in \partial S^2_{I} \cap \partial S^2_{II} \cap \partial S^2_{III}$, we will have $B^{\langle \hat{x}_1, \hat{x}_2,\hat{x}_5 \rangle}_{(\frac{1}{2},0,0,0,0)}$, $B^{\langle \hat{x}_1, \hat{x}_3,\hat{x}_5 \rangle}_{(\frac{1}{2},0,0,0,0)}$ and $B^{\langle \hat{x}_1, \hat{x}_4,\hat{x}_5 \rangle}_{(\frac{1}{2},0,0,0,0)}$ being excited, that is to say, the $3$ $W(S^2)$ operators generate a complex excitation with more complicated topology than $\beta$-chairon. But here the energy density along the excitation is uniform. Since the projection of the excitation onto a 2D plane has three ``petals'', this kind of excitations can be dubbed as ``cloverions''. Analogous to the $\beta$-chairon, we can obtain that there are $\binom{5}{1} \times \binom{4}{3}=20$ flavors of cloverions in the model-$[1,2,3,5]$.

\begin{figure*}[h]
	\centering  
	\subfigure[A $\beta$-chairon]{
		\label{fig_b_chair}
		\includegraphics[width=0.24\textwidth]{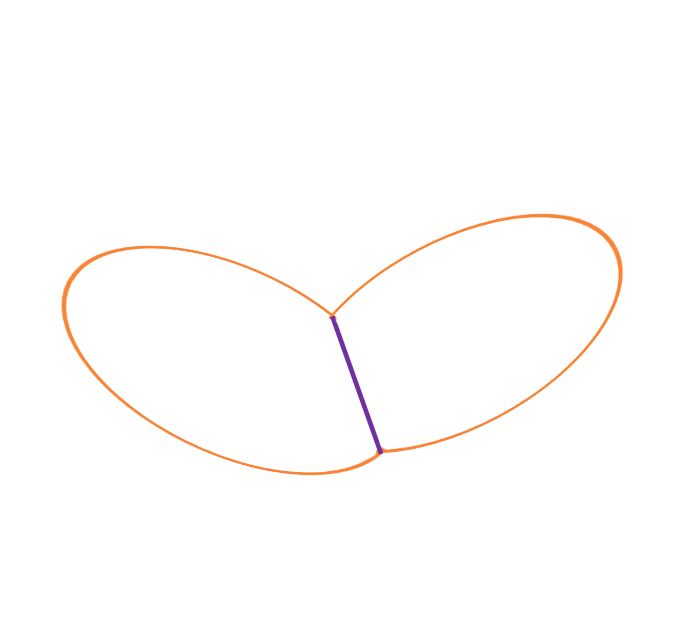}}
	\subfigure[A cloverion]{
		\label{fig_cloverion}
		\includegraphics[width=0.24\textwidth]{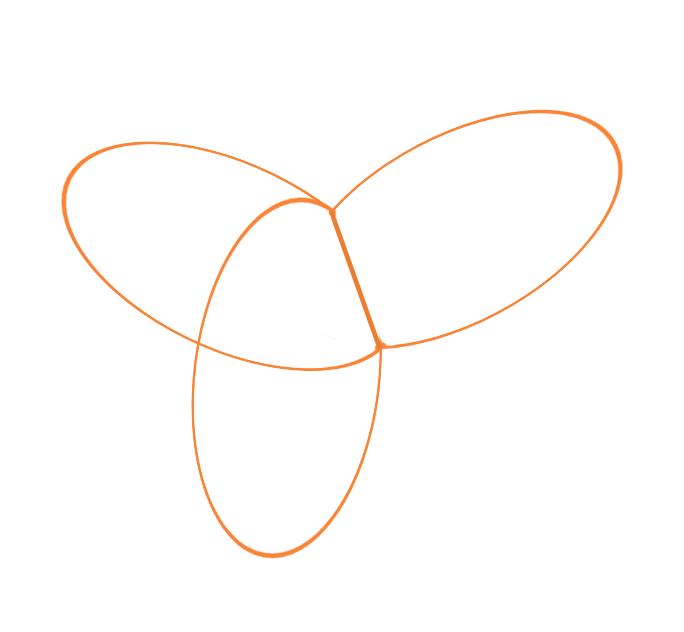}}
	\subfigure[A xuon]{
		\label{fig_xuon}
		\includegraphics[width=0.24\textwidth]{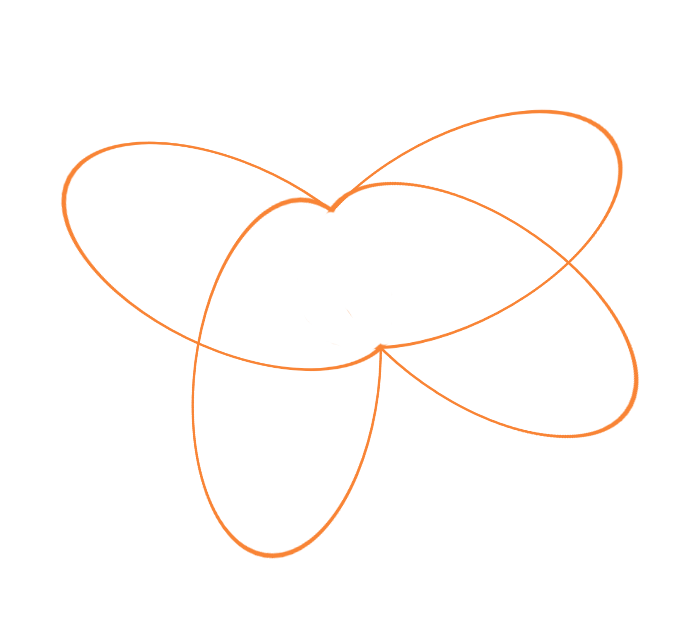}}
	\caption{  Pictorial comparison among a $\beta$-chairon, a cloverion and a xuon in the model-$[1,2,3,5]$. The crease line with higher energy density in $\beta$-chairon is highlighted with purple.}
	\label{fig_com_xuon}
\end{figure*}

But unlike the model-$[1,2,3,4]$, here we can further apply $W(S^2_{IV})$ to obtain an extra kinds of complex excitations, where $S^2_{IV}=\{(n+\frac{1}{2}, 0, 0, 0, m+\frac{1}{2})|m,n=0,1,2,\dots, L-1\}$. All $B$ operators associated with $(\frac{1}{2}, 0, 0, 0, 0)$ are unflipped by the $W(S^2_{IV})$ now, but four U-shaped strings generated by the four $W(S^2)$ operators will compose a complex excitation of a new kind. This kind of complex excitations can be dubbed as ``xuon'', as it is an X-shaped object consisted of 4 U-shaped strings. Obviously, there are only $\binom{5}{1}=5$ flavors of xuons in the model-$[1,2,3,5]$. A schematic comparison between $\beta$-chairons, cloverions and xuons is given in Fig.~\ref{fig_com_xuon}.

Except for $[1,2,3,5]$ and other two 5D models on the family tree (see Fig.~\ref{fig_tree})), we also have $[0,1,4,5]$, $[0,2,4,5]$ and $[0,3,4,5]$ models that are exactly solvable. Some representative excitations in more 5D models are listed in Table.~\ref{table:5d_excitations}. 

\begin{table*}[htp] 	
	\caption{Typical examples of excitations in more 5D models. Simple excitations are simply labeled by their mobility types in this table, so different excitations may share the same label. Names of other types of excitations are omitted for simplicity.}
	
	\begin{tabular}{ccccc}
		\hline
		
		\hline
		Models & Excitations &  Sectors & Flipped stabilizers & Creation operators \\
		\hline
		\multirow{2}{*}{$[0,1,4,5]$} & $(0,0)$ & $\mathsf{E}^s$ & $A_{\gamma_5}$ & $\prod\limits_{{\gamma}_1 \in D^4} \sigma^z_{{\gamma}_1}$ 
		\\
		& $(0,1)$ & $\mathsf{E}^s$ & $B^l_{\gamma_0}$ & $\prod\limits_{{\gamma}_1 \in S^1} \sigma^x_{{\gamma}_1}$
		\\
		&  & $\mathsf{E}^d$ & $A_{\gamma_5}$ & $\prod\limits_{{\gamma}_1 \in D^4} \sigma^z_{{\gamma}_1}$  
		\\
		\hline
		\multirow{2}{*}{$[0,2,4,5]$} & $(0,0)$ & $\mathsf{E}^s$ & $A_{\gamma_5}$ & $\prod\limits_{{\gamma}_2 \in D^3} \sigma^z_{{\gamma}_2}$ 
		\\
		& $(0,0)$ & $\mathsf{E}^s$ & $B^l_{\gamma_0}$ & $\prod\limits_{{\gamma}_2 \in S^2} \sigma^x_{{\gamma}_2}$
		\\
		& & $\mathsf{E}^d$ & $A_{\gamma_5}$ & $\prod\limits_{{\gamma}_2 \in D^3} \sigma^z_{{\gamma}_2}$  
		\\
		\hline
		\multirow{2}{*}{$[0,3,4,5]$} & $(0,0)$& $\mathsf{E}^s$ & $A_{\gamma_5}$ & $\prod\limits_{{\gamma}_3 \in D^2} \sigma^z_{{\gamma}_3}$ 
		\\
		& $(0,0)$ & $\mathsf{E}^s$ & $B^l_{\gamma_0}$ & $\prod\limits_{{\gamma}_3 \in S^3} \sigma^x_{{\gamma}_3}$
		\\
		&  & $\mathsf{E}^d$ & $A_{\gamma_5}$ & $\prod\limits_{{\gamma}_3 \in D^2} \sigma^z_{{\gamma}_3}$  
		\\
		\hline
		\multirow{2}{*}{$[2,3,4,5]$} & $(0,0)$ & $\mathsf{E}^s$ & $A_{\gamma_5}$ & $\prod\limits_{{\gamma}_3 \in D^2} \sigma^z_{{\gamma}_3}$ 
		\\
		& $(2,3)$ & $\mathsf{E}^s$ & $B^l_{\gamma_2}$ & $\prod\limits_{{\gamma}_3 \in S^3} \sigma^x_{{\gamma}_3}$
		\\
		&  & $\mathsf{E}^c$ & $B^l_{\gamma_2}$ & $\prod_{{\gamma_3}\in S^3_I} \sigma^x_{\gamma_3} \prod_{{\gamma_3}\in S^3_{II}} \sigma^x_{\gamma_3}$, where $\partial S^3_{I} \cap \partial S^3_{II} = C^2 \neq \emptyset$
		\\
		&  & $\mathsf{E}^d$ & $A_{\gamma_5}$ & $\prod\limits_{{\gamma}_3 \in D^2} \sigma^z_{{\gamma}_3}$ 
		\\
		
		\hline
		
		\hline	
	\end{tabular}
	\label{table:5d_excitations}
\end{table*}


\section{Concluding remarks}
\label{sec:conclusion}

In this article,   we've demonstrated that various kinds of novel excitations can be constructed in a large class of exactly solvable models of fracton topological order, like spatially extended excitations with restricted mobility and deformability, which unveils an intriguing scenario of interplay of topology and geometry in fracton order.   

 There are several future directions related to fracton physics of  spatially extended excitations.
 
 1. For instance, it is  worth to examine  more exactly solvable instances that are outside the family tree, like $[0,1,4,5]$, $[0,2,4,5]$, $[0,3,4,5]$ model in 5D, and other $12$ 6D models, and discuss their properties, e.g.,  exotic complex excitations,  fusion rules, entanglement entropy, and effective field theory. 
 
 2. As we've discussed in the main text, it is also interesting to explore the relation between geometry in fracton order and curved space caused by gravity.  
 
 3. By noting that volumeons denoted by $(0,3)$ can be constructed in some models of 4D or higher dimensions, one may conjecture that our   universe may have extra dimensions while elementary particles in the Standard Model are in fact volumeons that are actually restricted inside our 3D visible space. In this sense, it is very interesting to construct a higher dimensional lattice models that support volumeons which are massive Dirac fermions! Moreover, the relationship between the existence of complex excitations and the type of the order is also an exciting question. 
 
 4. Finally, it is also interesting to study self-localization theory of spatially extended excitations with different degrees of mobility and deformability restriction.

\acknowledgements
We thank Chenjie Wang, Meng Cheng, Han Ma, Hao Song, Juven  Wang, Andrey  Gromov, Kevin Slagle, Yizhi You, Jian-Keng Yuan, Zhi-Feng Zhang and Yuchen Ma for their  beneficial  communication and discussions. M.Y.L. \& P.Y. were supported in part by the Sun Yat-sen University startup grant and NSFC grant no. 11847608.  
  
   
%

\end{document}